\documentclass{aa}  

\usepackage{graphicx}
\usepackage{txfonts}
\usepackage[T1]{fontenc}

\usepackage{natbib}

\usepackage[colorlinks=true, linkcolor=blue, citecolor=blue, urlcolor=blue]{hyperref}

\begin{document}

\titlerunning{Redshift Evolution of Density Slopes}
\authorrunning{S. Geng et al.}

\title{Investigating the Redshift Evolution of Lensing Galaxy Density Slopes via Model-Independent Distance Ratios}

\author{S. Geng\inst{1}
    \and M. Grespan\inst{1}
    \and H. Thuruthipilly\inst{1}
    \and S. Harikumar\inst{1}
    \and A. Pollo\inst{1,2}
    \and M. Biesiada\inst{1}
}

\institute{National Centre for Nuclear Research, ul. Pasteura 7, 02-093 Warsaw, Poland\\
    \email{Shuaibo.Geng@ncbj.gov.pl}, \\
    \email{Margherita.Grespan@ncbj.gov.pl},\\
    \email{Hareesh.Thuruthipilly@ncbj.gov.pl}
    \and
    Astronomical Observatory of the Jagiellonian University, Faculty of Physics, Astronomy and Applied Computer Science, ul. Orla 171, 30-244 Kraków, Poland\\
}

\abstract
{
Strong lensing systems, expected to be abundantly discovered by next-generation surveys, provide a powerful tool for studying cosmology and galaxy evolution. The coupling between galaxy structure and cosmology through distance ratios makes examining the evolution of lensing galaxy mass density profiles essential for advancing both fields.
}
{
We introduce a novel, dark energy-model-independent method to investigate the mass density slopes of lensing galaxies and their redshift evolution using an extended power-law (EPL) model.
}
{
We adopt a non-parametric approach based on Artificial Neural Networks (ANNs) trained on Type Ia Supernovae (SNIa) data to reconstruct the distance ratios of strong lensing systems. These reconstructed ratios are compared with theoretical predictions to estimate the evolution of EPL model parameters. The analysis is performed at three levels - combined sample, individual lenses, and binned groups - to ensure robust estimates of the EPL parameters.
}
{
A negative evolutionary trend of mass density power-law exponent with increasing redshift is observed across different analysis levels. Assuming a triangular prior for the anisotropy of lensing galaxies, we find evidence for redshift evolution of the mass density slope, quantified as $\partial\gamma/\partial z = -0.20 \pm 0.12$.
}
{
This study confirms that the redshift evolution of matter density slopes in lensing galaxies can be determined independent of dark energy models at the population level. Using simulations informed by  Legacy Survey of Space and Time (LSST) Rubin Observatory forecasts, which is expected to identify 100,000 strongly lensed galaxies, we show that spectroscopic follow up of just 10\% of these systems can reduce the uncertainty in the redshift evolution coefficient of the total mass density slope ($\Delta\partial\gamma/\partial z$) to 0.021. Such precision would be able to distinguish between evolving and non-evolving pictures for lensing galaxies.
}

\maketitle

\section{Introduction} \label{sec:intro}
Despite its limitations, the $\Lambda$ Cold Dark Matter ($\Lambda$CDM) model remains the standard paradigm for describing the Universe at the largest scales. Various well-established cosmological probes have been used within this framework to constrain the cosmological parameters. These probes encompass a wide range of techniques, including cosmic microwave background (CMB) observations \citep{planckcollaboration2020planck}, baryon acoustic oscillations (BAO) \citep{eisenstein2005detection,cole20052df,desicollaboration2024desi} and supernovae Ia (SNIa) \citep{Riess_SN, Perlmutter,riess2021comprehensive}.
Moreover, alternative approaches using strong gravitational lenses (SGLs) \citep{suyu2014cosmology,cao2015cosmology,arendse2019lowredshift,liu2019implications,liao2019modelindependent,wong2020h0licowa,liao2020determining,qi2021measurements,li2024cosmology} and gravitational waves \citep{abbott2017gravitationalwave,gayathri2020hubble} are being used as complementary to the above methods. However, statistically significant discrepancies are observed between the results obtained from these various methods, commonly referred to as the Hubble tension \citep{divalentino2021realm} and the S8 tension \citep{asgari2021kids1000,harnois-deraps2024kids1000}. For a detailed discussion on cosmological tensions, please refer to \citet{divalentino2021realm}. They highlight the need for further exploration and refinement of both observational techniques (including careful study of possible systematics and biases) and cosmological theories. 
 
As a particularly valuable cosmological probe, SGL can provide independent constraints on cosmological parameters. Rooted in the theory of general relativity (GR), SGL describes the deflection of light from a distant source by the gravitational field of a massive intervening object. This phenomenon offers unique information about the link between the structure of galaxies and cosmological information, making it a powerful tool for studying both cosmological parameters and galaxy evolution.  

A key observable in studies of SGLs is the position and shape distortion of lensed images, which can be obtained with high-resolution data. Such information allows us to determine the Einstein radius $\theta_E$.
This quantity represents the angular radius of the ring-like image formed when light from a perfectly aligned background source is bent by the gravity of lensing galaxies. $\theta_E$ provides a robust measurement of the total projected mass of the lensing object. Furthermore, when combined with stellar kinematics data, it can help to constrain the radial mass profile of the lens \citep{treu2002internal,koopmans2003structure,koopmans2006gravitational}. However, the observed angular size of the Einstein radius depends on the cosmological model, as it is influenced by the distances between the source, lens, and observer, specifically via the distance ratio ${D^A_{ls}}/{D^A_s}$, 
where $D^A_{ls}$ is the angular diameter distance from the lens to the source, and $D^A_s$ is the angular diameter distance from the source to the observer.
The dependence mentioned above makes the Einstein radius a valuable observable for testing cosmological models, provided there is an accurate lens model for comparison with theoretical predictions. Such an approach has already been employed in testing dark energy models \citep{cao2011testing, cao2012testing}, testing cosmic curvature \citep{liu2020testinga,wang2022cosmological}, constraints of the cosmological models \citep{Biesiada2010,cao2012constraintsa,cao2015cosmology}. 

One of the prerequisites in the approaches mentioned above is an assumption about the mass profile of the lensing galaxy. In this work, we use the extended power-law (EPL) model, where power-law density profiles describe both luminous and dark matter. Allowing for different radial density distributions between the luminous and dark matter components provides the freedom to describe the mass distribution of the lensing galaxy in a more realistic way. The EPL model was developed by \citet{koopmans2006gravitational}, and it provides an analytical solution that enhances computational efficiency, leading to its broad usage in various applications. These include constraining cosmological parameters \citep{chen2019assessing,li2024cosmology}, exploring the Friedmann-Lemaître-Robertson-Walker (FLRW) metric \citep{cao2019direct,wang2022cosmological}, testing theories of general relativity \citep{lian2022direct, guerrini2024probing}, examining dark matter models \citep{bora2021probing,cheng2021new}, and studying galaxy evolution \citep{geng2021velocity}.

Is the EPL model adequate for statistical analysis of SGL samples? Previous cosmological model-dependent studies of individual lensing galaxies have indicated a slight redshift evolution in the radial total mass density slope \citep{sonnenfeld2013sl2sb}, a trend also noted in broader statistical studies \citep{cao2015cosmology,cao2016limits,holanda2017constraintsa}. Moreover, \citet{chen2019assessing} highlighted the significance of incorporating redshift-dependent evolution of the mass density power-law exponent and the intrinsic scatter of the luminous density slope on cosmological outcomes. They showed that accounting for such evolution in the mass models of lensing galaxies is needed to improve cosmological constraints. 
The evolution of mass distribution with redshift also offers insight into the merger history of elliptical galaxies \citep{remus2013dark,tan2024joint}. In particular, dissipative, gas-rich mergers tend to contract dark matter halos, steepening the total density slope of early-type galaxies at lower redshifts. Conversely, dissipation-less, gas-poor mergers cause halo expansion, resulting in a shallower density slope with decreasing redshift. Developing general and robust approach to account for potential redshift evolution in mass density distributions is the most important objective of this work. It is essential for enhancing population-level cosmological constraints and advancing our understanding of galaxy evolution. This need is particularly pronounced in the context of upcoming large-scale surveys, such as the Legacy Survey of Space and Time (LSST) from the Vera C. Rubin Observatory \citep{ivezic2019lsst} and the Euclid telescope\citep{acevedobarroso2024euclid} which are expected to increase the number of observed strong lensing events significantly.

In this paper, we assume a spatially flat and transparent Universe, wherein the luminosity distance $D^L$ and the angular diameter distance $D^A$ adhere to the cosmic distance duality relation ${D^L(z)}/{D^A(z)}=(1+z)^2$. 
To achieve a model-independent\footnote{In this paper, `model-independent' refers to the approach of not assuming any fixed background cosmology like $\Lambda$CDM model with definite (fiducial) values of the Hubble constant, matter density parameter, and dark energy density parameters. It should be better called a `dark-energy' model agnostic approach. Let us point out that in the literature the term of `model-independent' approach is still being used, while it is almost impossible to avoid prior cosmological assumptions, like the (flat) FLRW model or Etherington principle (i.e. the distance duality relation), as we spelled out explicitly in the text.} reconstruction of the redshift-angular diameter distance relationship $D^A(z)$, we use Artificial Neural Networks (ANNs) trained on type Ia supernovae (SNIa) observations. The distance ratios reconstructed purely from the data are then compared with theoretical predictions to evaluate the parameters of the EPL model and assess their redshift evolution in the sample of lensing galaxies.

The structure of the paper is as follows. Section \ref{sec:theory} sets the theoretical framework for our lens model. Section \ref{sec:data} outlines the data used, including those from SNIa, along with the SGLs data. In Section \ref{sec:Methodology}, we detail our approach for distance reconstruction from observations and the statistical methods used to determine density slopes and their evolution with redshift. Section \ref{sec:results} presents our results, followed by discussion and conclusions in Section \ref{sec:discussions} and Section \ref{sec:summary}, respectively.

\section{Theoretical background}\label{sec:theory}

This section provides a concise overview of the parametric lens model employed in our analysis and details the method used to constrain the density slopes for the lensing galaxies.

The Singular Isothermal Sphere (SIS) model is a widely employed mass model for investigating galaxy-galaxy lensing. 
 This model assumes a spherically symmetric mass distribution for the lensing galaxy, with a density profile,  $\rho(r)$ that scales inversely with the square of the radial distance, i.e. $\rho(r)\propto r^{-2}$. Due to this spherical symmetry, the SIS model predicts the occurrence of two images.  
However, real galaxies often exhibit elliptical shapes and produce quadruple-imaged sources. To account for this, the Singular Isothermal Ellipsoid \citep[SIE;][]{Kormann1994} model introduces ellipticity into the lens potential. 

This is achieved by substituting  $r \rightarrow \sqrt{q_m x^2+{y^2}/{q_m}}$, where $q_m$ is the axis ratio, and the coordinates $(x,y)$ are defined by rotating the (RA, Dec) coordinate system by the position angle of the semi-major axis so that $x$ coincides with this axis. In such models, the Einstein radius is defined by 
\begin{equation} \label{eq:Einstein radius}
\theta_E=4\pi\frac{\sigma_{v,SIS}^2}{c^2}\frac{D^A_{ls}}{D^A_s}.
\end{equation}
Here, $\sigma_{v,SIS}$ represents the stellar velocity dispersion of the lensing galaxy being the SIS model parameter. We can estimate this quantity spectroscopically by measuring the velocity dispersion within an aperture $\sigma_{ap}$.
This value needs to be rescaled to approximate the central velocity dispersion $\sigma_{v,e/2}$, which refers to the velocity dispersion at the half-effective radius of the galaxy. Since these two quantities may not coincide in practice, a nuisance parameter $f_E$ correcting for possible biases $\sigma_{v,SIS} = f_E \sigma_{v,e/2}$ was introduced in many previous research \citep[e.g. in][]{cao2012constraintsa}.

The projected mass $M_{Ein}$ within the Einstein radius is given by 
\begin{equation} \label{eq:lensing mass}
M_{Ein}=\pi \theta_E^2 \left(D^A_l\right)^2\Sigma_{cr},
\end{equation}
where $D^A_l$ is the angular diameter distance from the lens to the observer, and $\Sigma_{cr}=({c^2}/{4\pi G})\cdot {D^A_s}/(D^A_lD^A_{ls})$ denotes the critical projected mass density which is the threshold for strong lensing. 
Although the simple SIS/SIE models are sufficient for many practical purposes, they assume a prior radial mass distribution. One should use more sophisticated models to constrain the radial dependence of the mass profile in the lensing galaxy.    

Considering both baryonic and dark matter components, which significantly contribute to the total mass distribution of the galaxy, we adopt the extended power-law mass density model from \citet{koopmans2006gravitational}. 
Without losing simplicity, this model assumes that the luminous tracer and the total mass in the lensing galaxy follow power-law distributions in the radial direction, represented as $\rho_{lum}(r)$ for luminous mass and $\rho_{tot}(r)$ for the total mass:
\begin{equation} \label{eq:mass distribution assumption}
\begin{aligned}
&\rho_{lum}(r)=  \rho_{lum,0}r^{-\delta}\\
&\rho_{tot}(r)= \rho_{tot,0}r^{-\gamma}\\
&\beta(r)=1-\frac{\langle\sigma_{v,\theta}^2\rangle}{\langle\sigma_{v,r}^2\rangle},
\end{aligned}
\end{equation}
where $\delta$ and $\gamma$ are the logarithmic density slopes of the luminous baryonic matter and the total matter (dark matter plus luminous matter), respectively. The parameter $\beta$ describes the anisotropy of the three-dimensional velocity dispersion of the luminous tracer within the lensing galaxy. It depends on the ratio of the tangential velocity dispersion $\sigma_{v,\theta}$ and the radial velocity dispersion $\sigma_{v,r}$. Upon solving the spherical Jeans equation under these assumptions, one can derive the dynamical mass $M_{dyn}$ of the lensing galaxy \citep{koopmans2006gravitational}, which corresponds to the projected mass enclosed within $\theta_E$:
\begin{equation} \label{eq:dynamical mass}
M_{dyn} = \frac{\pi}{G} \sigma_{v}^2 D^A_l \theta_E \left( \frac{\theta_E}{\theta_{ap}} \right)^{2-\gamma} f(\gamma,\delta,\beta),
\end{equation}
where
\begin{equation} \label{eq:general f function}
\begin{aligned}
f(\gamma,\delta,\beta)&= \frac{1}{2\sqrt{\pi}}\left(\frac{(\gamma+\delta-5)(\gamma+\delta-2-2\beta)}{\delta-3} \right) \\
&\times \frac{\Gamma\left( \frac{\gamma+\delta-2}{2}\right)\Gamma\left( \frac{\gamma+\delta}{2}\right)}{\Gamma\left( \frac{\gamma+\delta}{2}\right)\Gamma\left( \frac{\gamma+\delta-3}{2}\right)-\beta \Gamma\left( \frac{\gamma+\delta-2}{2}\right)\Gamma\left( \frac{\gamma+\delta-1}{2}\right)} \\
&\times  \frac{\Gamma\left( \frac{\delta-1}{2}\right)\Gamma\left( \frac{\gamma-1}{2}\right)}{\Gamma\left( \frac{\delta}{2}\right)\Gamma\left( \frac{\gamma}{2}\right)}.
\end{aligned}
\end{equation}
Here $\Gamma$ denotes the Gamma function. The formula above displays the mass inferred from the luminosity-weighted average line-of-sight velocity dispersion $\sigma_{v}$ within the specified aperture $\theta_{ap}$ of the spectrograph and illustrates the usefulness of the power-law profile to scale between $\theta_E$ and $\theta_{ap}$. In practice, constant $\theta_{ap}$ would correspond to different physical scales in different galaxies, hence, as it will be discussed in details later one adjusts the velocity dispersion to the half of the effective radius of given galaxy.
The value of $\beta = 0$ indicates an isotropic, spherically symmetric density distribution, and setting both $\gamma$ and $\delta$ to 2 recovers the SIS model. By assuming that the dynamical mass (Eq.~\ref{eq:dynamical mass}) equals the gravitational mass within the Einstein radius (Eq.~\ref{eq:lensing mass}), we can define a theoretical distance ratio as:
\begin{equation} \label{eq:theoretical distance ratio}
\begin{aligned}
\mathcal{D}^{th}({\hat{p}}_{free};{\hat{p}}_{given})= & \frac{D^A_{ls}}{D^A_s} \\ 
=& \frac{c^2}{4 \pi } \frac{\theta_E}{\sigma_{v}^2} \left( \frac{\theta_E}{\theta_{ap}} \right)^{\gamma - 2} f^{-1}(\gamma),
\end{aligned}
\end{equation}
where ${\hat{p}}_{free}$ and ${\hat{p}}_{given}$ are free parameters and given (i.e. measured) parameters, respectively. 

It is worth noting that in a spatially flat Universe, the distance ratio is independent of the Hubble constant $H_0$. 
In a transparent Universe, the distance sum rule allows us to express the observed distance ratio in terms of comoving distances $D(z)$ or luminosity distances $D^L(z)$:
\begin{equation} \label{obeserved dd}
\mathcal{D}^{obs}=\frac{D^A_{ls}}{D^A_s}= 1 - \frac{D_l}{D_s}=
1-\frac{1+z_S}{1+z_L}\frac{D^L_l}{D^L_s}.
\end{equation}
Knowing the redshifts $z_L$ and $z_S$, the corresponding luminosity and comoving distances can be determined from the distance-redshift relations reconstructed purely from the SNIa. These methods will be detailed in the next section. By comparing $\mathcal{D}^{obs}$ with the theoretical distance ratio $\mathcal{D}^{th}({\hat{p}}_{free};{\hat{p}}_{given})$ predicted by the chosen lens model, we can infer the density slope $\gamma$ and other model parameters. This can be achieved by minimizing the $\chi^2$ objective function:
\begin{equation} \label{eq:Chi square}
    \chi^2({\hat{p}}_{free};{\hat{p}}_{given}) = \sum_i \frac{ \left[ \mathcal{D}_i^{obs} - \mathcal{D}_i^{th}\right]^2}{(\Delta \mathcal{D}_i^{obs})^2 + (\Delta \mathcal{D}_i^{th})^2} .
\end{equation}
By virtue of the uncertainty propagation formula the absolute uncertainty\footnote{In order to avoid confusion with the velocity dispersion we denote the uncertainty by $\Delta$ instead of commonly used $\sigma$.} of $\mathcal{D}^{obs}$ is given by
\begin{equation} \label{obs uncertainty}
\Delta \mathcal{D}^{obs}=\frac{1+z_S}{1+z_L}\frac{\sqrt{\left(\Delta D^L_l\right)^2 + \left(\Delta D^L_s\right)^2 \left( D^L_l/D^L_s\right)^2}}{D^L_s},
\end{equation}
and the absolute uncertainty of $\mathcal{D}^{th}$ is given by
\begin{equation} \label{theo uncertainty}
\Delta\mathcal{D}^{th}=\mathcal{D}^{th}\sqrt{4(\delta\sigma_{v})^2+(1-\gamma)^2(\delta\theta_E)^2},
\end{equation}
where the fractional uncertainties $\delta\sigma_{v}$ and $\delta\theta_E$ will be discussed in Section \ref{sec:data}. In our analysis, we disregard the covariance between velocity dispersion and Einstein radii, as these measurements are derived from different instruments and methodologies. For simplicity, we assume that there are no correlated uncertainties between them.

\section{Data} \label{sec:data}

\subsection{Type Ia supernovae}\label{sec:Ia}
Type Ia supernovae are high-energy explosions originating from the white dwarfs within binary stellar system \citep{Hoyle,Colgate}. They hold significant importance in cosmology as they can be used as standard candles \citep{Riess_SN,Perlmutter}. The luminosity distance $D_L(z)$ to a SN Ia can be expressed as  
\begin{equation} \label{distanc_sn}
    \mu (z) = 5\mathrm{log_{10}} \left(\frac{D_L(z)}{\mathrm{Mpc}}\right) + 25,
\end{equation}
where $\mu (z)$ is the observed distance modulus of the SN Ia.

To train our ANNs (more details below), we used 1550 unique, spectroscopically confirmed SNe Ia from 18 different surveys (Pantheon+ data set) compiled by \citet{Scolnic}. Their publicly available SNIa sample\footnote{\url{https://github.com/PantheonPlusSH0ES/DataRelease}}  provides the measured $\mu(z)$ and $z$. The sample covers a wide redshift span from $z=0.0012$ to $z=2.2617$, which makes it one of the largest samples for cosmological studies. Using the measured $\mu (z)$ value, we estimated the luminosity distance to the supernovae using Eq.~\ref{distanc_sn}, which is then converted to the angular distance to train our ANN. The $\Delta \mu$ values in the Pantheon+ dataset are used to estimate distance uncertainties. The ANN outputs are fixed with a shape of (batch size, 2), representing the distance and its error. Since the ANN cannot directly output a covariance matrix, we did not include the Pantheon+ covariance matrix in our analysis. As a result, the ANN predicts larger errors than the actual measurement uncertainties. Properly incorporating the covariance matrix could improve distance predictions which would need better deep learning models. However, this study serves as a proof of concept, demonstrating that distance ratios from deep learning models can predict the mass-density slope. Improving the model is planned for future work.

\subsection{Strong gravitational lenses sample}\label{subsec:SGL_sample}
In this work, we use a galaxy-scale SGLs sample, originally assembled by \citet{cao2015cosmology} and later updated by \citet{chen2019assessing}. This sample comprises 161 galaxies, predominantly early-type (E/S0 morphologies), carefully selected to exclude those with significant substructures or proximate companions. The dataset integrates 5 systems from the LSD survey \citep{koopmans2002stellar, koopmans2003structure, treu2002internal, treu2004massive}, 26 from the Strong Lenses in the Legacy Survey (SL2S)  \citep{ruff2011sl2s, sonnenfeld2013sl2sa, sonnenfeld2013sl2sb, sonnenfeld2015sl2s}, 57 from the Sloan Lens ACS Survey (SLACS) \citep{bolton2008sloan, auger2009sloan, auger2010sloanb}, 38 from the an extension of the SLACS survey known as “SLACS for the Masses (S4TM)” \citep{shu2015sloan, shu2017sloan}, 21 from the BELLS \citep{brownstein2012boss}, and 14 from the BELLS for GALaxy-Ly EmitteR sYstems (GALLERY) \citep{shu2016boss, shu2016bossa}. 

\begin{figure}
\centering
\includegraphics[width=\linewidth]{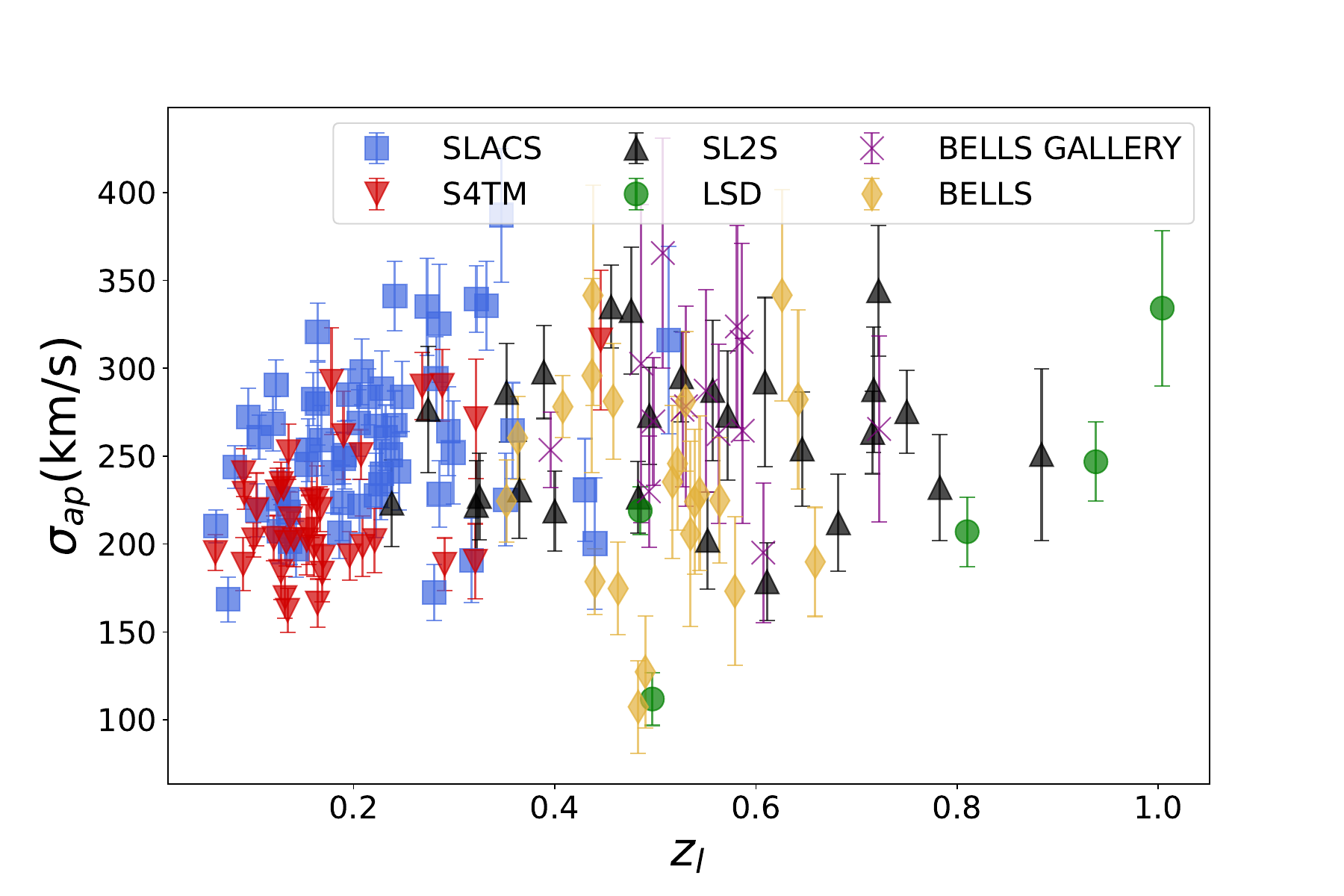}
\caption{\label{fig:scatter}Distribution of corrected velocity dispersions and redshifts for lensing galaxies in the combined sample from various surveys which are used in this work. Blue squares represent lenses from SL2S, green dots represent lenses from BELLS, red down triangles represent lenses from BELLS GALLERY, purple crosses represent lenses from SLACS, black triangles represent lenses from LSD, and yellow diamonds represent lenses from S4TM. }
\end{figure}

For our analysis, we used the following observables: (1) spectroscopic redshifts of the lensing galaxies ($z_L$) and the sources ($z_S$); (2) the Einstein radius ($\theta_E$), with an assumed uncertainty of $\delta\theta_E=5\%$, (3) the effective radius ($\theta_{\text{eff}}$), i.e. the half-light radius for a galaxy; and (4) the velocity dispersion ($\sigma_{ap}$) within an aperture of angular radius ($\theta_{ap}$) along with its uncertainty. The velocity dispersion is measured using a rectangular slit, hence the equivalent circular aperture is calculated as  \citep{jorgensen1995spectroscopy}:
\begin{equation} \label{eq:effective aperture}
\theta_{ap}\simeq 1.025\times\sqrt{(\theta_x\theta_y/\pi)}.
\end{equation}
To ensure consistency across different physical sizes measured in the fix-sized aperture, we normalized measured velocity dispersions to 
the apertures equivalent to half of the effective radius $\theta_{e/2}$, denoted by $\sigma_{v,e/2}$, employing the aperture correction formula from \citet{jorgensen1995spectroscopy}:
\begin{equation} \label{eq:aperture correction}
\sigma_v^{\text{obs}} = \sigma_{e/ 2} = \sigma_{\text{ap}} \left( \frac{\theta_{\text{eff}}}{2\theta_{\text{ap}}} \right)^\eta.
\end{equation}
This particular scale was chosen because it closely aligns with the most probable scale of the Einstein radius. 
For the value of the correction parameter, we adopt $\eta=-0.066 \pm 0.035$, as determined by \citet{cappellari2006sauron}. Following \citet{chen2019assessing}, the total uncertainty in the velocity dispersion is calculated as:
\begin{equation} \label{eq:VD uncertainty}
(\Delta \sigma_{e/2}^{\text{tot}})^2 = (\Delta \sigma_{e/2}^{\text{stat}})^2 + (\Delta \sigma_{e/2}^{\text{AC}})^2 + (\Delta \sigma_{e/2}^{\text{sys}})^2,
\end{equation}
where $\Delta \sigma_{e/2}^{\text{stat}}$ represents the statistical uncertainty from the measurements, $\Delta \sigma_{e/2}^{\text{AC}}$ is propagated uncertainty from the aperture correction to half effective radius, which is calculated as
\begin{equation} \label{eq:aperture correction uncertainty}
\begin{aligned}
(\Delta \sigma_{e/2}^{\text{AC}})^2 &= (\Delta \sigma_{ap})^2 \left(\frac{\theta_{eff}}{2 \theta_{ap}}\right)^{2 \eta} \\
&+  \sigma_{ap}^2 \left(\frac{\theta_{eff}}{2 \theta_{ap}}\right)^{2 \eta} \left( \ln{\frac{\theta_{eff}}{2 \theta_{ap}}}\right)^2 (\Delta \eta)^2,
\end{aligned}
\end{equation}
and $\Delta \sigma_{e/2}^{\text{sys}}$ accounts for the systematic uncertainty associated with the assumptions made regarding the consistency between $M_{dyn}$ and $M_{Ein}$. We incorporate a systematic uncertainty of 3\% on the model-predicted velocity dispersion to account for the potential influence of line-of-sight matter on the Einstein mass, as suggested by \citet{jiang2007baryon}. 
From the overall uncertainty mentioned in Eq.~\ref{eq:VD uncertainty}, we can calculate the relative uncertainty $\delta\sigma_{ap}$ in Eq.~\ref{theo uncertainty} by $\delta\sigma_{ap} = \Delta \sigma_{e/2}^{\text{tot}} / \sigma_{e/2}$ .
The corrected velocity dispersion and the redshift distribution of our combined sample are shown in Fig.~\ref{fig:scatter}.

\subsection{Simulated Dataset}\label{sec:simulation}
We generated synthetic datasets of SN Ia redshifts and galaxy-galaxy strong lensing observables based on LSST forecast models. This serves two main purposes: (a) to assess whether our pipeline can accurately recover the ground truth, and (b) to predict the constraining power of future wide-field surveys like LSST. 
The LSST main survey is expected to detect approximately 400,000 photometrically classified SNe Ia and around 100,000 strong lensing systems \citep{collett2015population,ivezic2019lsst}. For our simulations, we assume a flat $\Lambda$CDM cosmology with $\Omega_{M}=0.3$ and $h=0.7$, consistent with the simulated strong lens population of \citet{collett2015population}.

\begin{figure}
\centering
\includegraphics[width=8.5cm]{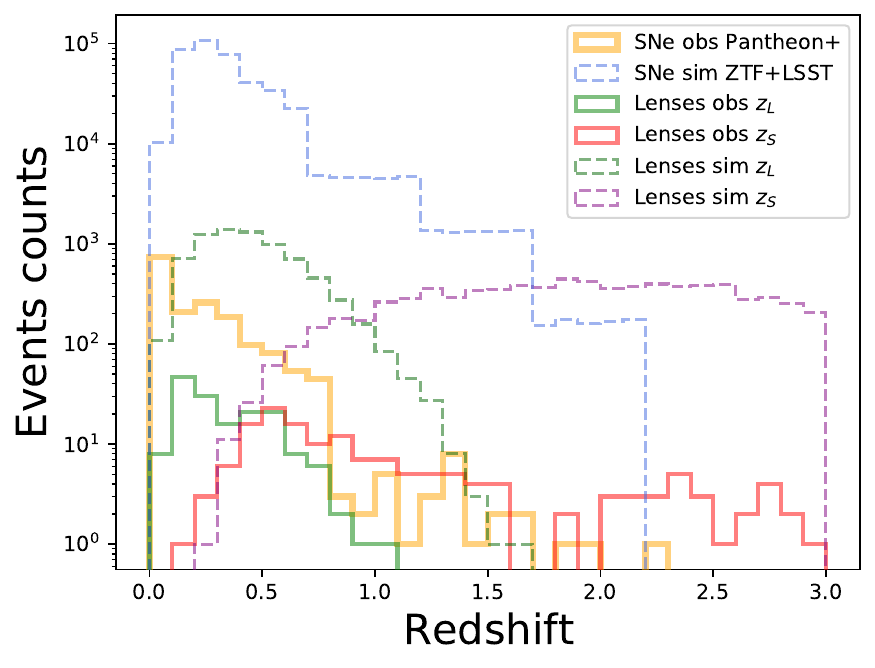}
\caption{The redshift distributions for the simulated and observed SNe Ia, lenses, and background sources are presented. "obs" represents the observed populations from current observations, while "sim" denotes the mock populations simulated in this study.
\label{fig:redshift hist}}
\end{figure}

The simulated SNe Ia dataset was divided into two groups: low-redshift ($z<0.16$) and high-redshift. For the low-redshift group, we generate 12,000 SNe Ia with spectroscopic redshifts,  scaling up from the redshift distribution of current Zwicky Transient Facility (ZTF) observations, which recorded over 3,000 SNe Ia in 2.5 years \citep{dhawan2022zwicky}. For the high-redshift sample, we simulated 400,000 SNe Ia with spectroscopic redshifts, adopting the Pantheon+ redshift distribution. In the high-redshift group, redshift bins of 0.05 were used for $0.1 < z \leq 0.7$, and bins of 0.5 were applied for $z > 0.7$. 
For both redshift groups, redshifts were randomly assigned in each bin and luminosity distances were computed based on the fiducial cosmology described above. We included a 10\% relative uncertainty in the luminosity distance to reflect the typical observational uncertainties in current SN Ia measurements. Figure~\ref{fig:redshift hist} shows the resulting redshift distribution of the mock SNe Ia.

To model the lens population expected from LSST, we followed the approach of \citet{li2024cosmology}, along with the lensing population forecast from \citet{collett2015population}. Assuming a 10\% spectroscopic confirmation rate for strong lens candidates, we obtained a sample of 7,506 systems with source redshifts below 3.0, consistent with spectroscopic galaxy-galaxy lens catalogs. From the SGL population, we took lens redshifts, source redshifts, and Einstein radii, which we used to calculate the velocity dispersion. During this process, we incorporated probability distributions for the slopes of the mass density ($\gamma$ and $\delta$) and the anisotropy parameter ($\beta$). 
Specifically, we adopted $\gamma = 2.078 \pm 0.16$ \citep{auger2010sloanb} and $\delta = 2.173 \pm 0.085$ \citep{chen2019assessing}, while setting $\beta = 0.22 \pm 0.2$, as discussed in \ref{subsubsec:beta_prior}.
In our real-data SGL catalog described in \ref{subsec:SGL_sample}, we found an average relative error of 11\% for $\Delta \sigma_{e/2}^{\text{tot}}/ \sigma_{ap}$. Hence, we applied this relative uncertainty to the mock velocity dispersion values. No redshift evolution in lens galaxy properties was assumed. The resulting redshift distributions of the simulated lens and background source populations are also shown in Fig.~\ref{fig:redshift hist}.

\section{Methodology} \label{sec:Methodology}
This section details our methodology for distance reconstruction and constraining the density slope of lensing galaxies, including its evolution with redshift. For reproducibility, the code used for this work is publicly available on GitHub \footnote{\url{https://github.com/margres/SGL_gamma}}.

\subsection{Distance reconstruction through ANN} \label{sec:dd}
Distance measurements are a critical aspect of SGL analysis, as they are intricately linked to both the observed data and the underlying cosmological model.
SGL modeling tools often rely on a fiducial model, such as the `vanilla' $\Lambda$CDM cosmology, to calculate angular diameter distances. SGL systems have proven to be an effective cosmological probe independent of the distance ladder, enabling the determination of key parameters such as the matter energy density $\Omega_m$ and the cosmic equation of state parameter $w$, as highlighted in studies by \citet{chae2007cosmological,oguri2008sloan,Biesiada2010,cao2015cosmology, liu2019implications}. 
Thus, the ideal scenario would be to measure this quantity independently of any specific cosmological model. This would provide a direct measurement, making it less susceptible to biases inherent in cosmology-dependent methods. In this work, we aim to derive distances directly from objects at corresponding redshifts, assuming only a flat and transparent universe.

Artificial Neural Networks are powerful computational tools inspired by the structure and function of the human brain. They excel at identifying complex patterns within data. An ANN is built from interconnected processing units called neurons, arranged in layers. Each neuron applies a non-linear mathematical function (activation function) to its input, and the outputs are then transmitted to neurons in subsequent layers.
The training process of an ANN involves adjusting the weights associated with these connections to optimize the performance of the network. This optimization minimizes an error function (loss function), essentially enabling the network to learn from the provided data.
The Universal Approximation Theorem by \citep{HORNIK1990551}
states that an ANN with at least one hidden layer containing a finite number of neurons can approximate any continuous function, given a continuous and non-linear activation function. This remarkable capability allows ANNs to learn the underlying relationships within complex datasets, like cosmological data, without any pre-defined model of the functional form.
ANNs have been shown to be effective in reconstructing the relation between redshift and luminosity distance for SN Ia \citep{Gomez, Dialektopoulos_ann_2023, Shah_2024_ann, Mitra_2024_ann} as well as for other cosmological probes such as cosmic chronometers \citep{Gomez, Mukherjee_2022}.

In this work, we implemented a 3-layer ANN with 20 neurons in each layer.
 The network took the redshift of the SNIa as input and was trained to predict the corresponding angular diameter distance along with the associated error.
 During training, the learning rate was set to 0.001. We used the Exponential Linear Unit \citep[ELU,][]{Clevert} activation function for all layers, except the final output layer, as it helps speed up learning in deep neural networks . Additionally, ELU improves classification accuracy compared to commonly used activation functions like the Rectified Linear Unit \citep[ReLU,][]{agarap2018deep} function. The hyperparameters of the model, including its structure, were determined through a random grid search.
To initialize the network weights, we employed the Xavier uniform initializer \citep{Glorot}. The entire network was trained from scratch using the  ADAM optimizer with its default exponential decay rates \citep{kingma2017adam}. 
To prevent overfitting, we implemented an early stopping callback from {\tt Keras} \footnote{\url{https://keras.io/api/callbacks}} that monitors the validation loss and automatically halts training when the validation loss stops decreasing.

\subsection{Constraints on logarithmic density slope with the single slope model (\texorpdfstring{$\gamma=\delta, \beta=0$}{gamma=delta, beta=0})}

We commence our analysis with the simplest scenario: a single slope lens model. This model assumes no distinction between the total and luminous mass distribution, resulting in a single logarithmic density distribution, i.e. $\gamma=\delta$.
To estimate the posterior distribution of the model parameters, we employ  Markov chain Monte Carlo (MCMC) techniques using the Python module {\tt emcee} \citep{foreman-mackey2013emcee}.
This approach allows us to perform log-likelihood sampling based on the $\chi^2$ function defined in Eq.~\ref{eq:Chi square}.
To mitigate potential biases originating from the specific SGL data used in the MCMC analysis, we perform our investigation using various methodologies.
First, we examine the posteriors obtained from individual lenses (Section \ref{sec:meth_sing}). Second, we analyze lenses binned by their scale factor and redshift values with a fixed step size (Section \ref{sec:binning}). Finally, we consider the posteriors derived from the entire lens sample (Section \ref{sec:direct}). This multi-faceted  analysis strategy helps to ensure the robustness of our conclusions and mitigate the influence that might arise from the characteristics of individual lenses or specific groups of lenses.

We extract the values of the logarithmic density distribution from the posterior distribution. The median of the posterior distribution represents the most probable value for $\gamma$, while the associated uncertainty is estimated using the median absolute deviation (MAD). The MAD is defined as:
\begin{equation}
\mathrm{MAD}=\mathrm{median}(|X_i-\mathrm{median}(X)|).
\end{equation}
After obtaining logarithmic density slopes, $\gamma$ values from the SGL analysis at different distances, we proceed to extract their redshift dependence. This is achieved by fitting a linear regression to the relationship between  $\gamma$ and $z$.

\subsubsection{Individual constraints}\label{sec:meth_sing}
For each SGL system, indexed by $i$, the theoretical distance ratio is calculated as $\mathcal{D}^{th}_i(\gamma_i;[z_L,z_S,\sigma_{obs},\theta_E,\theta_{eff}]_i)$ from observables. 
In this approach, we use MCMC sampling to minimize the $\chi^2$ statistic, defined in Eq.~\ref{eq:Chi square}, for each lensing system individually. 
The $\chi^2$ for the $i$-th individual lensing system is given by ${ \left( \mathcal{D}_i^{obs} - \mathcal{D}_i^{th}\right)^2}/\left[ (\Delta \mathcal{D}_i^{obs})^2 + (\Delta \mathcal{D}_i^{th})^2\right]$. From the resulting posterior distribution, we derive the median and the MAD of the density slope parameter $\gamma_i$ for each system.
In order to investigate the redshift evolution of the density slope, we fit a linear relationship ($\rm\gamma^{Sin-I} = \gamma_{0}^{Sin-I}+\gamma_{s}^{Sin-I}\times z_L$) on obtained $\gamma_i$ values across different lens redshifts $z_{l,i}$.

\subsubsection{Fixed-bin approach}\label{sec:binning}
In contrast to the individual lens analysis detailed in Section \ref{sec:meth_sing}, this method focuses on minimizing Eq.~\ref{eq:Chi square} for a group of lenses. We achieve this by binning the lenses according to two distinct strategies: redshift and scale factor ($a$).

\begin{figure}
\centering
\includegraphics[width=8.5cm]{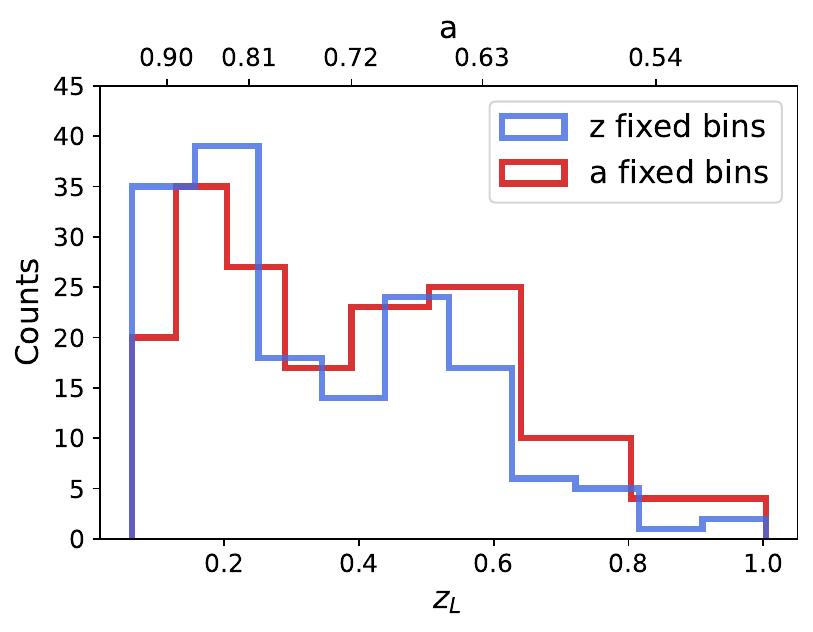}
\caption{Distribution of lens' redshifts for the full sample. The blue histogram shows bins with fixed redshift intervals, while the red histogram shows bins with fixed scale-factor sizes.
\label{fig:bin hist}}
\end{figure}

The first strategy involves binning lenses based on constant redshift intervals. We adopt a bin size of $dz$=0.1 spanning the entire redshift range of our lens sample. This approach allows for a straightforward grouping of lenses with similar redshifts, potentially revealing trends in the data.
However, as redshift and distance do not have a linear relationship, we also consider a binning strategy based on the scale factor $a = {1}/(1+z)$. We set a constant step size for this binning, calculated as $da = (a_{max} - a_{min})/8$ for each subsample, ensuring the distribution captures significant data trends, particularly in lower scale factor bins, which are supposed to contain at least 5 data points. The distributions of redshifts and scale factors within the bins for the full sample are shown in Fig.~\ref{fig:bin hist}. We can find that the scale factor bins contain more SGL system at high redshift end.

Assuming uniform logarithmic density distribution for both total and luminous mass, the theoretical distance ratio for systems in each bin is represented as $\mathcal{D}^{th}_i(\gamma^{\rm Sin-bin}_{\alpha};[z_L,z_S,\sigma_{obs},\theta_E,\theta_{eff}]_i)$, where the subscript $i$ denotes the $i$-th lensing system within either the $\alpha$-th fixed-size redshift bin or the $\alpha$-th fixed-size scale factor bin. Similar to the approach outlined in Section \ref{sec:meth_sing}, we address the redshift evolution of the density slope by conducting a linear fit to the median values and MADs of the posterior distribution of $\gamma$ from each bin. 

\subsubsection{Full sample ``direct'' fit} \label{sec:direct}

This approach deviates from the previously described methods by directly incorporating a redshift evolution of the mass density slope within the MCMC analysis applied to the entire SGL sample. This allows us to infer the redshift dependence of the mass density slope $\gamma$ without the need for an intermediate step involving linear fits to the individual $\gamma$ values obtained from the posterior distribution sampling (Sections~\ref{sec:meth_sing} and \ref{sec:binning}).

We still assume the same mass distribution for the total and luminous components. However, we consider a linear relationship between the total mass logarithmic slope and redshift, $\rm\gamma_i^{Sin-d} = \gamma_{0}^{Sin-d}+\gamma_{s}^{Sin-d}\times z_{l,i}$, as a universal characteristic of the sample. This affects the theoretical distance ratio as $\mathcal{D}^{th}_i(\gamma_{0}^{\rm Sin-d},\gamma_{s}^{\rm Sin-d};[z_L,z_S,\sigma_{obs},\theta_E,\theta_{eff}]_i)$ for the $i$-th lensing system.

\begin{figure}
\centering
\includegraphics[width=8.5cm]{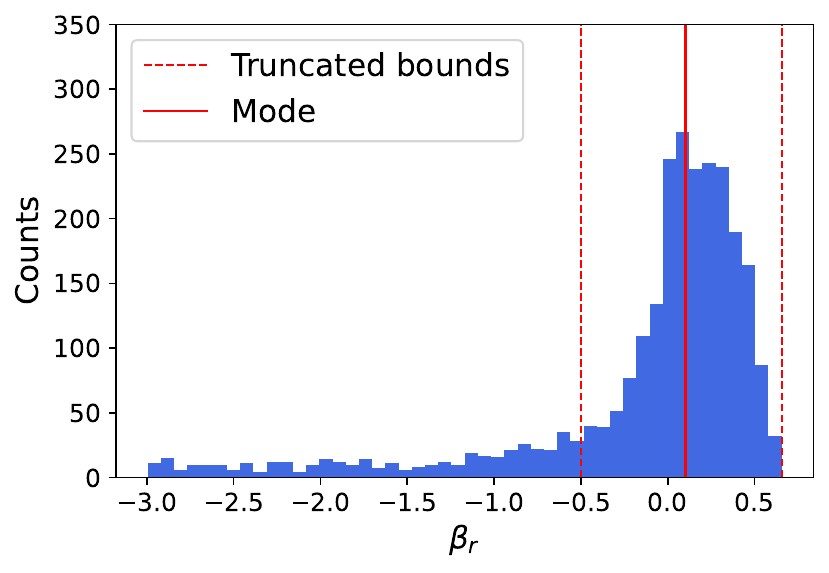}\caption{Distribution of the velocity anisotropy constraints for early-type galaxies from stellar dynamic modeling for the MaNGA survey in the final SDSS DR17 \citep{zhu2023manga}.
\label{fig:beta_r}}
\end{figure}

\subsection{Constraints on logarithmic density slope with extended spherical power-law model (\texorpdfstring{$\gamma \neq \delta$}{gamma != delta})}

In this method, we differentiate between the logarithmic density slopes for total ($\gamma$) and luminous ($\delta$) matter in Eq.~\ref{eq:mass distribution assumption}. The anisotropy ($\beta$) of the velocity dispersion of the lensing galaxy is also included as a free parameter with a prior. We consider two scenarios for these density slopes: non-evolving and evolving with redshift. 

\subsubsection{Lensing galaxy velocity dispersions anisotropy prior} \label{subsubsec:beta_prior}
The dynamical anisotropy of velocity dispersions within lensing galaxies poses a challenge for accurate modeling. To address this challenge, we explore the impact of different prior distributions for $\beta$ on our MCMC analysis. We consider three distinct priors:

1. A commonly used truncated Gaussian prior, based on 40 nearby elliptical galaxies from SAURON IFS data \citep{jorgensen1995spectroscopy,bolton2006constraint}, uses $\beta = {\cal N}(0.18;0.13)$ and is truncated within the range [$\beta - 2\sigma_\beta$, $\beta + 2\sigma_\beta$].

2. We introduce a new prior utilizing Dynamics and Stellar Population analysis from the MaNGA Survey \citep[SDSS Data Release 17,][]{zhu2023manga,zhu2024manga}.  
The MaNGA Survey utilizes integral field unit (IFU) data to provide spatially resolved dynamical information for nearly $10^4$  local universe galaxies, analyzed via the axisymmetric Jeans Anisotropic Modeling (JAM) method. 
To ensure compatibility with our spherical power-law model, we refine velocity anisotropy constraints for early-type galaxies using the $\mathrm{JAM_{sph}}$ model assuming the general NFW dark matter distribution outlined by \citet{zhu2023manga}. Galaxy morphology is sourced from the deep learning-based catalog by \citet{dominguezsanchez2022sdssiv}. To focus on reliable data, we selected only early-type galaxies with a visual quality rating greater than 0 from \citet{zhu2023manga},  resulting in a sample of 2597 galaxies. The distribution of their velocity anisotropy values is shown in Fig.~\ref{fig:beta_r}. From this dataset, we identify a peak value (mode) of $\beta_p$ at 0.102. We establish a triangular prior distribution centered at mode, with bounds at $(-0.5,0.656)$, encompassing the majority of galaxies in the sample.

3. Based on the same MaNGA sample described above, \citet{guerrini2024probing} implemented an additional selection criterion. They focused on galaxies where the absolute difference between the velocity anisotropy values obtained from the NFW model (NFW) and the generalized NFW model (gNFW) was less than 0.05 ($|\beta_{\mathrm{NFW}}-\beta_{\mathrm{gNFW}}|<0.05$).
Applying this criterion to the original MaNGA sample \citep{zhu2023manga} resulted in a refined selection of 1136 galaxies. The fit of a truncated logistic Gaussian distribution yielded a best-fit prior of $\beta = {\cal N}(0.22;0.2)$ truncated at $[\mu-3\sigma,\mu+3\sigma]$.

\subsubsection{Non-evolving lens model}
The non-evolving scenario assumes that our SGL systems can be effectively described by the extended power-law model with some constant values of $\gamma$ and $\delta$ representative of the whole sample. The theoretical distance ratio for each lensing system is represented as $\mathcal{D}^{th}_i(\gamma^{\rm EPL-n},\delta^{\rm EPL-n},\beta;[z_L,z_S,\sigma_{obs},\theta_E,\theta_{eff}]_i)$, where the anisotropy parameter $\beta$ is not a free parameter, but is represented by a certain prior as discussed in Sect. \ref{subsubsec:beta_prior}. 

\subsubsection{Evolving lens model}\label{sec:evo_binning}
Similar to the single density slope model, we implement different approaches to explore potential redshift evolution features. Here, only fixed-bin approach and the direct approach are applied, since the bad performance of the three-parameter constraint on individual SGL system. For fixed-bin approach, we independently determine the density slopes in each bin without accounting for redshift evolution, subsequently fitting these constraints linearly across the bins to identify any evolutionary trends. The direct approach adopts two semi-analytical models for redshift evolution. The first model is linear evolutions for $\gamma$ and $\delta$ with respect to the lens redshift $z_L$, represented as
\begin{equation} \label{eq:linear z evolution}
\begin{aligned}
\gamma^{\rm EPL-lin} = \gamma_{0}^{{\rm EPL-lin}}+\gamma_{s}^{{\rm EPL-lin}}\times z_L; \\
\delta^{\rm EPL-lin} = \delta_{0}^{{\rm EPL-lin}}+\delta_{s}^{{\rm EPL-lin}}\times z_L,
\end{aligned}
\end{equation}
where the superfix `$\rm EPL-lin$' denotes the linear evolution parameters under the extended spherical power-law model. \\
In order to check the evolution with respect to the scale factor, we consider the second evolving model as
\begin{equation} \label{eq:CPL z evolution}
\begin{aligned}
\gamma^{\rm EPL-ATE} = \gamma_{0}^{{\rm EPL-ATE}}+\gamma_{s}^{{\rm EPL-ATE}}\times \frac{z_L}{1+z_L}; \\
\delta^{\rm EPL-ATE} = \delta_{0}^{{\rm EPL-ATE}}+\delta_{s}^{{\rm EPL-ATE}}\times \frac{z_L}{1+z_L},
\end{aligned}
\end{equation}
where the superfix `$\rm EPL-ATE$' denotes the evolution model comes from the Taylor expansion in terms of scale factor. In order to compare the performance of different $\beta$ priors and evolving scenarios, we employ corrected Akaike Information Criterion (cAIC) and Bayesian Information Criterion (BIC) evaluations. They are, respectively, defined as
\begin{equation} \label{eq:AICBIC}
\begin{aligned}
&\mathrm{cAIC}=\frac{2k(k+1)}{N-k-1}+2k+\chi_{\mathrm{min}}^2; \\
&\mathrm{BIC}=k\ln{N}+\chi_{\mathrm{min}}^2,
\end{aligned}
\end{equation}
where k is the number of parameters and N is the number of data points. BIC assesses model quality by approximating $-2\ln{(\mathrm{marginal\; likelihood})}$ with fitting the maximum likelihood estimation. Additionally, it adds a penalty of the complexity based on the number of parameters. The cAIC is also commonly used for model selection and it adjust for small sample size data based on the standard AIC. The cAIC only differs from the BIC in its penalty term.

\section{Results} \label{sec:results}
\subsection{Distance reconstruction}
\begin{figure}
\centering
\includegraphics[width=8.5cm]{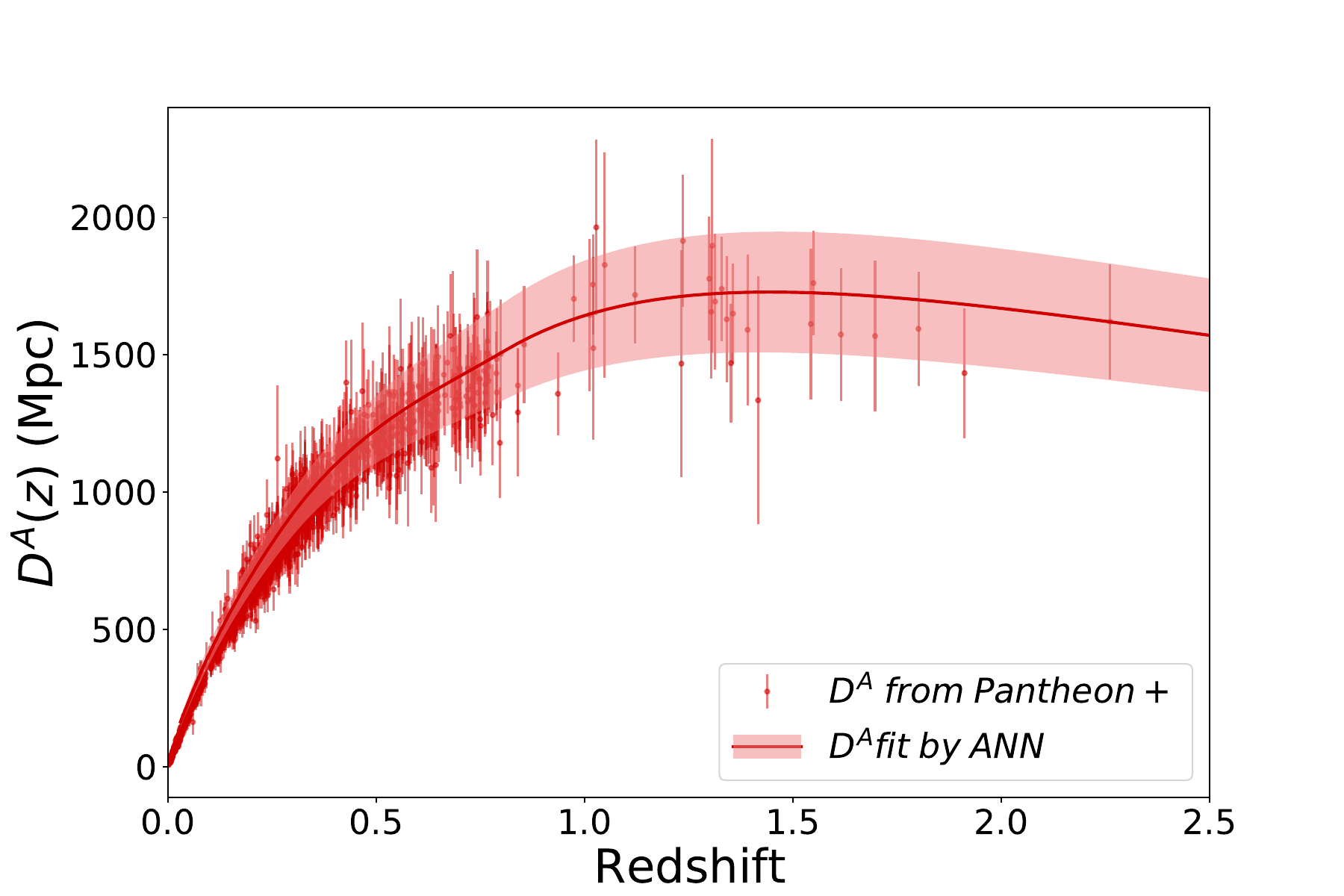}
\caption{\label{fig:dis reconstruction}Reconstruction of angular diameter distance under non-parametric methods. Red dots indicate the angular diameter distance calculated from the measurements of SNIa. Red line represents the best-fitted distance reconstruction by ANN+SNIa. The shadow region shows the 1 $\sigma$ uncertainties for corresponding distance reconstruction.}
\end{figure}

We employed the model-independent ANN method, which performed without assuming any specific cosmology parameters, just a flat transparent $\Lambda$CDM structure, to reconstruct the angular diameter distance at various redshifts.
The resulting distance-redshift relationships are visualized in Fig.~\ref{fig:dis reconstruction}. Despite a limited number of high-redshift data points, the reconstructed distance-redshift relation accurately captures the expected feature of the angular diameter distance, notably showing an inflection point around redshift 1.5. These reconstructed distances serve as the foundation for calculating the observed distance ratio $\mathcal{D}^{obs}$ (see Eq.~\ref{obeserved dd}) for each lensing system within our analysis.

\subsection{Constraints from single density slope model}

\begin{figure}
\includegraphics[width=8.5cm]{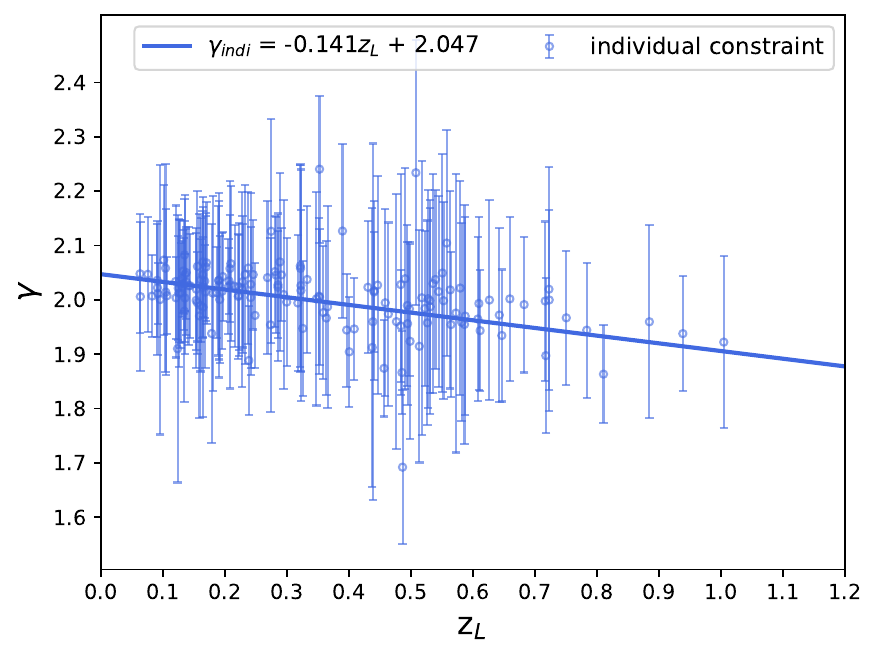}
\caption{The linear fit for the individual constraints on logarithmic slope $\gamma$.
\label{fig:singular linear fit}}
\end{figure}

In this section, we present the constraints on the redshift evolution of the density slope $\gamma$ of single uniform density distributions using the individual approach, the binning approach, and the direct method, respectively. The best-fitted values for $\gamma_0^{\rm Sin}$ and slopes $\gamma_s^{\rm Sin}$ are listed in the Table~\ref{tab:sigular and direct}. The results from the individual method suggest that $\gamma$ decreases with increasing redshift, as shown by ${\partial \gamma^{\rm Sin-I}}/{\partial z_L} = -0.141 \pm 0.033$, displayed in Figure~\ref{fig:singular linear fit}.

\begin{figure}
\includegraphics[width=8.5cm]{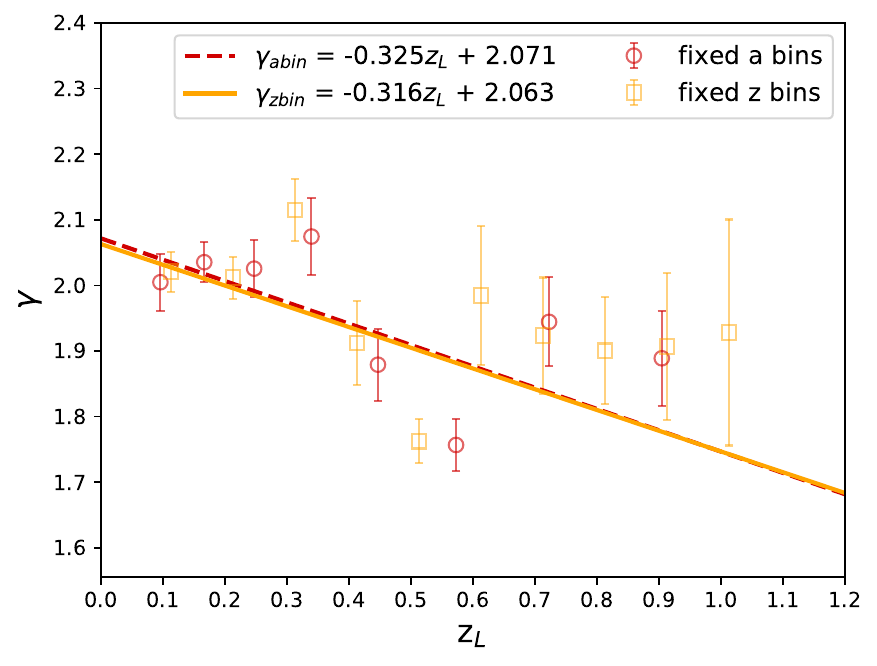}
\caption{The linear fit for $\gamma$ constrained in fixed redshift bins. Red hollow circles represent results in fixed scale factor bins, while orange hollow squares denote results in fixed redshift bins.
\label{fig:binning}}
\end{figure}

Analyzed within fixed-size bins, the best-fit linear redshift evolution of $\gamma$ is depicted in Fig.~\ref{fig:binning}. Within the fixed-size redshift bins, the combined sample indicates a redshift evolution for the density slope $\gamma^{\rm Sin-zbin}=2.063(\pm0.015)-z_L\times 0.316(\pm0.053)$. For the fixed-size scale factor bins, the best-fitted linear redshift evolution of $\gamma$ presents $\gamma^{\rm Sin-abin}=2.070(\pm0.013)-z_L\times 0.311(\pm0.036)$. The results of different binning approaches agree well with each other. 
As we can see in Fig.~\ref{fig:binning}, due to limited observations of redshifts close to 1, uncertainty increases in these bins.

In the direct constraining approach, assuming global redshift evolution, the results yield $\gamma^{Sin-d}=2.069(\pm0.028)-z_L\times 0.294(\pm0.073)$. Despite the considerable uncertainties for each SGL system in the individual approach, it shows a lower uncertainty compared to other approaches. Compared to the individual approach, the binning method and the direct approach suggest steeper redshift evolution for lensing galaxies.

\begin{table*}[ht]
\caption{The summary of the constraints on the intercept ($\rm\gamma_0$) and the slope ($\rm\gamma_s={\partial \gamma}/{\partial z_L}$) of the linear redshift evolution across different approaches.}
\centering
    \label{tab:sigular and direct}
    \renewcommand{\arraystretch}{1.5}
    \begin{tabular}{l|c|c|c|c}
        \hline \hline
        Parameters & Individual & Fixed z bins & Fixed a(z) bins & Direct \\
        \hline
        $\rm \gamma_0^{Sin}$                        & 2.047 $\pm$ 0.011  & 2.063 $\pm$ 0.015  & 2.071 $\pm$ 0.013  & 2.069 $\pm$ 0.028 \\
        $\rm\gamma^{Sin}_s$ & -0.141 $\pm$ 0.033 & -0.316 $\pm$ 0.053 & -0.325 $\pm$ 0.036 & -0.294 $\pm$ 0.073 \\
        \hline
        $\rm \gamma_0^{EPL}$   (Triangular prior)   & - & 2.129$\pm$0.030  & 2.104$\pm$0.038  & 2.065 $\pm$ 0.046 \\
        $\rm\gamma^{EPL}_s$   (Triangular prior)    & - & -0.400$\pm$0.073 & -0.32$\pm$0.10   & -0.20 $\pm$ 0.12 \\
        \hline \hline
    \end{tabular}
\end{table*}

\subsection{Constraints with \texorpdfstring{$\gamma \neq \delta$ model}{gamma != delta}}

\subsubsection{Non-evolving \texorpdfstring{$\gamma$ and $\delta$}{gamma and delta}}\label{res:non-evo}

\begin{figure}
\centering
\includegraphics[width=9.0cm]{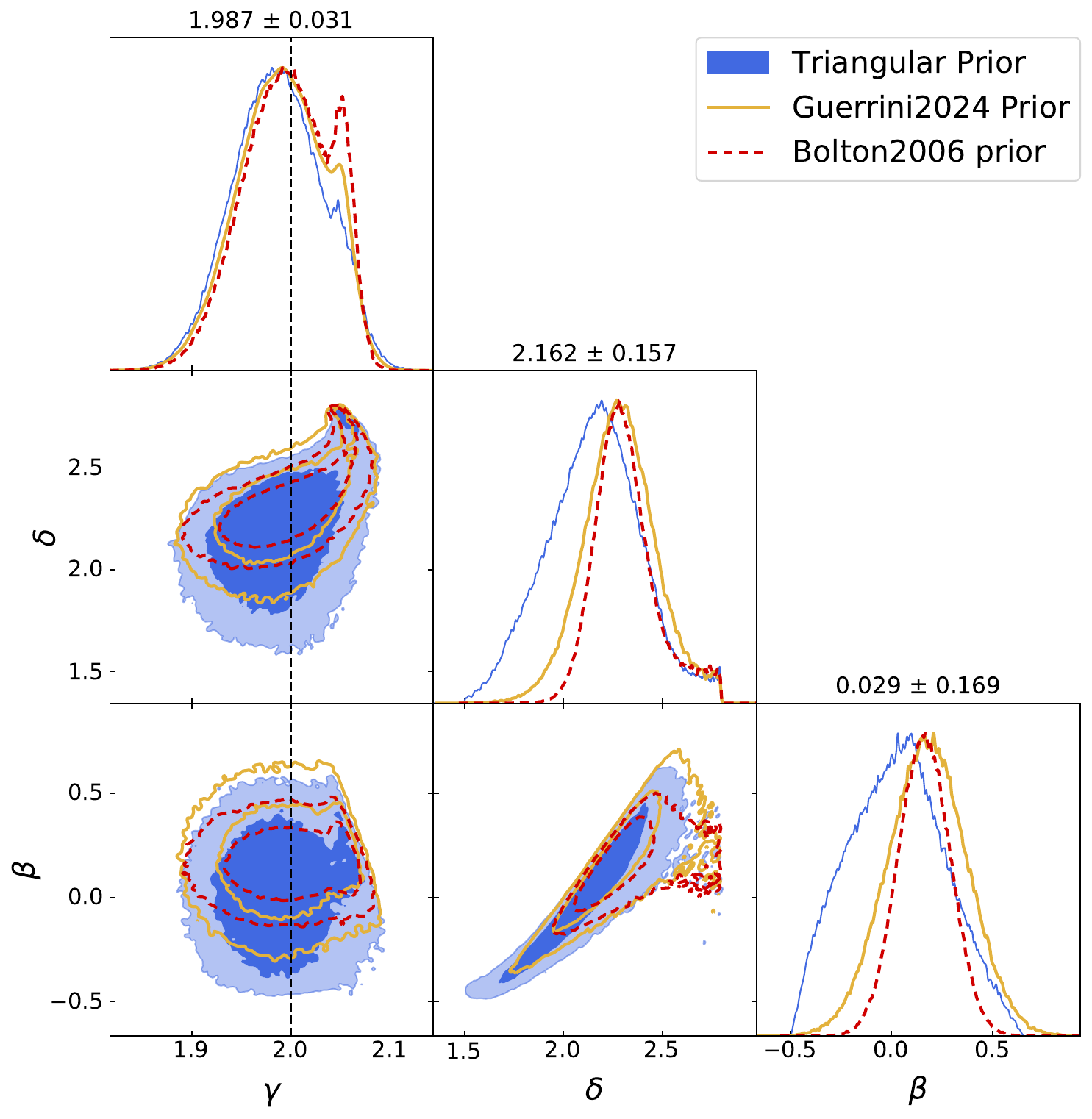}
\caption{\label{fig:sph3D}The constraints on the logarithmic density slope of the total mass and the luminous mass in the non-evolution picture. The blue contours indicate the posterior distribution adopting a triangular prior of $\beta$ defined in Section \ref{subsubsec:beta_prior}. The yellow and red contours indicate the posterior distribution adopting the Gaussian priors of $\beta$ based on \citet{guerrini2024probing} and \citet{bolton2006constraint}, respectively. The black dashed line corresponds to the singular isothermal sphere.}
\end{figure}

Assuming a static mass model, we constrain the density slopes of total mass and luminous matter using different anisotropy priors, which are described in Section ~\ref{subsubsec:beta_prior}. For the median values and MAD of the posterior distributions: The triangular prior yields a total mass density slope of $\gamma^{\rm EPL-n}=1.987\pm0.031$ and a luminous density slope of $\delta^{\rm EPL-n}=2.16\pm0.16$. A Gaussian prior, suggested by \citet{guerrini2024probing}, indicates $\gamma^{\rm EPL-n}=1.993\pm0.032$ and $\delta^{\rm EPL-n}=2.29\pm0.12$. Another Gaussian prior from \citet{bolton2006constraint} results in $\gamma^{\rm EPL-n}=1.998\pm0.032$ and $\delta^{\rm EPL-n}=2.298\pm0.092$. As depicted in Fig \ref{fig:sph3D}, a strong degeneracy exists between the velocity anisotropy parameter $\beta$ and the luminous matter density slope $\delta$, highlighting the necessity of selecting a reasonable $\beta$ prior. A narrower range for $\beta$ provides more stringent constraints on $\delta$ but does not significantly affect the constraining power on $\gamma$. The spherical model, without accounting for redshift evolution, displays a bimodal distribution in $\gamma$ constraints, indicating possible variations in the total mass density slope of lensing galaxies that may be related to other quantities such as redshift or velocity dispersions.

\subsubsection{Evolving \texorpdfstring{$\gamma$ and $\delta$}{gamma and delta}}

With the full sample, we firstly utilize the binning approach to constrain the density slope in the redshift bins and scale factor bins, respectively, and fit the linear redshift evolution. As shown in Table~\ref{tab:evo_3D}, the total mass density slopes from different $\beta$ priors show a strong negative redshift evolution, and they do not have significant deviations. While the constraint on $\delta$ evolution varies with the change of $\beta$ priors.

\begin{figure}
\centering
\includegraphics[width=\linewidth]{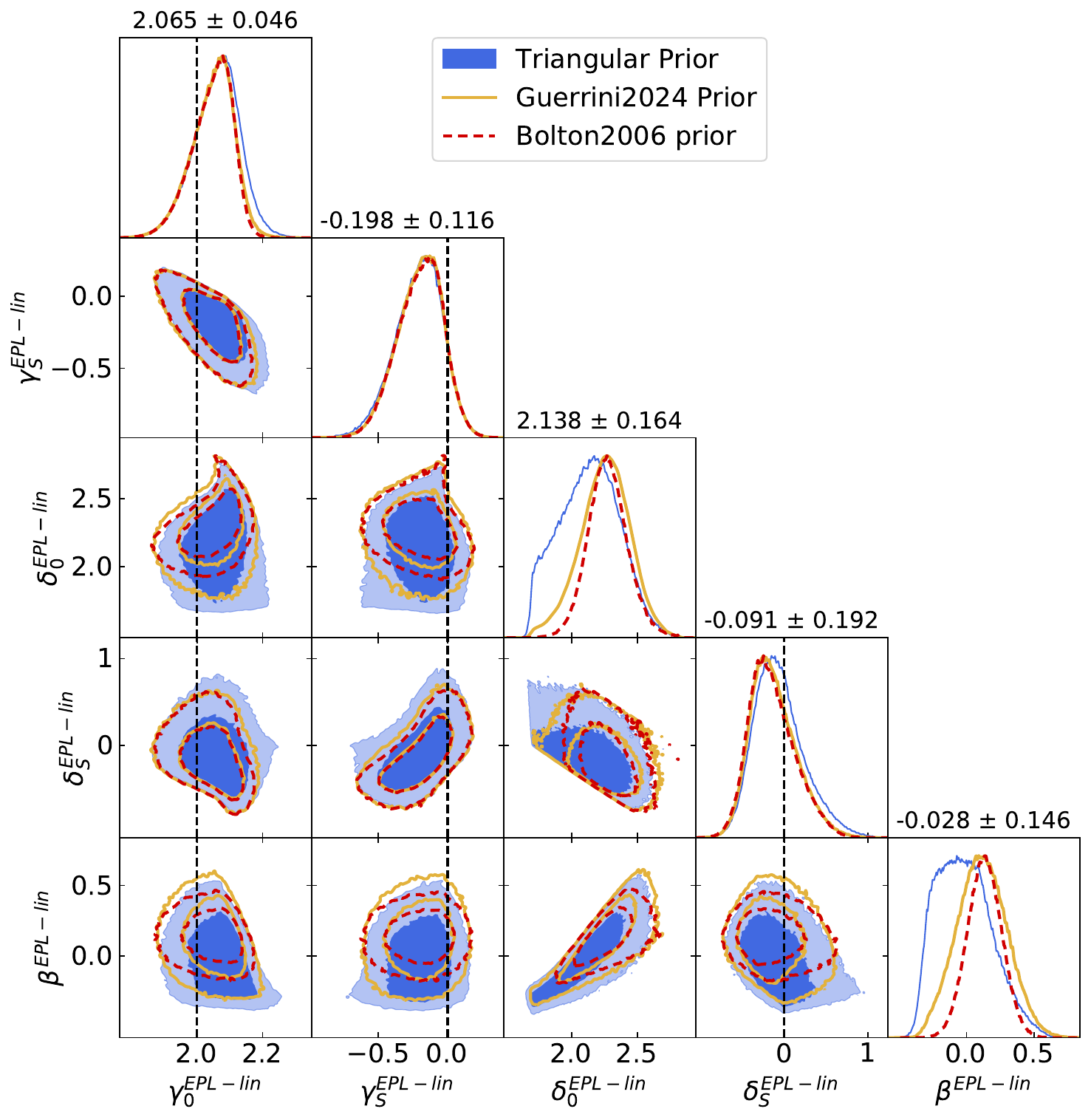}
\caption{\label{fig:linear z evo}Posterior distributions of the spherical power-law density model parameters with linear redshift evolution. The blue contours indicate the posterior distribution adopting a triangular prior of $\beta$ defined in Section \ref{subsubsec:beta_prior}. The yellow and red contours indicate the posterior distribution adopting the Gaussian priors of $\beta$ based on \citet{guerrini2024probing} and \citet{bolton2006constraint}, respectively. The black dashed line corresponds to the singular isothermal sphere.}
\end{figure}

Next, we apply linear redshift evolution models to the mass density slopes directly. In order to ensure the precision of our constraints, we select lensing systems where the relative uncertainty in the normalized velocity dispersion ($\Delta\sigma_{e/2}^{\mathrm{tot}}/\sigma_{v,e/2}$) is less than $20\%$. This criterion restricts our sample to 132 lensing systems with well-constrained velocity dispersions, thus it can alleviate the potential systemic errors in determining the mass density slope. The constraints from the triangular prior for the linearly evolving spherical model show a median total mass density slope ($\gamma_0$) of $2.065\pm0.046$ and a redshift evolution gradient ($\gamma_s$) of $-0.20\pm0.12$. The luminous mass density slope intercept ($\delta_0$) is $2.14\pm0.16$, with a gradient ($\delta_s$) of $-0.09\pm0.19$. The results in Fig.~\ref{fig:linear z evo} indicate a slight negative redshift evolution for both total and luminous mass densities, without ruling out a non-evolving scenario for luminous matter within 1$\sigma$ uncertainty. Also, the result confirms that the luminous mass density slope is steeper than that of total mass. Strong degeneracies are observed between $\delta_0$ and $\beta$. 
Comparing results from different $\beta$ priors, Table~\ref{tab:evo_5D} shows that the wider range triangular prior matches the constraining strength of the narrower Gaussian priors for $\gamma$ evolution constraints.

At last we apply the ATE evolution model to the mass density slopes with three $\beta$ priors. The triangular prior yields a total mass density slope redshift evolution as $\gamma^{EPL-ATE}=2.079(\pm0.056)-0.35(\pm0.23)\times{z}/{(1+z)}$, and the luminous mass density slope redshift evolution $\delta^{EPL-ATE}=2.17(\pm0.17)-0.25(\pm0.40)\times{z}/{(1+z)}$. 
In contrast to linear evolution, the ATE redshift evolution model indicates a steeper density slope of total mass at lower redshifts and aligns more closely with isothermal behavior at higher redshifts. Again the triangular prior is preferred by cAIC and BIC under the ATE evolution model, but is surpassed by the linear evolution model with triangular prior. Across different parametric models, all constraints show a negative redshift evolution in the total mass density slope of lensing galaxies. Moreover, the 1$\sigma$ uncertainty region of the best-fit models from both linear and ATE parameterization encompasses the isothermal scenario. 

\begin{figure}
\centering
\includegraphics[width=\linewidth]{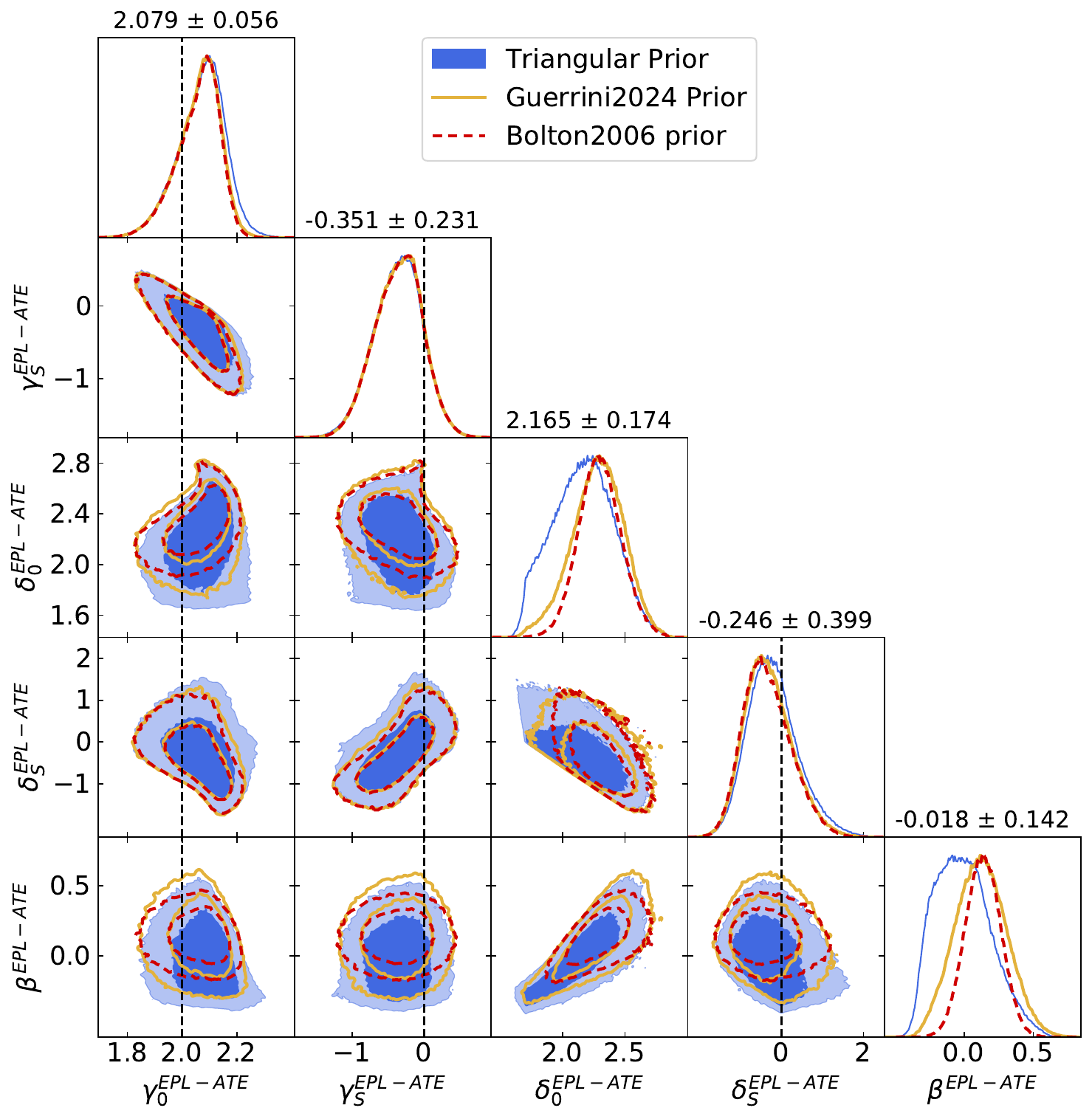}
\caption{\label{fig:ATE evo}Posterior distributions of the spherical power-law density model parameters under ATE evolution. The blue contours indicate the posterior distribution adopting a triangular prior of $\beta$ (see Section \ref{subsubsec:beta_prior}). The yellow and red dashed contours represent the posterior distribution derived from Gaussian priors on $\beta$ following\citet{guerrini2024probing} and \citet{bolton2006constraint}, respectively. The black dashed line corresponds to the singular isothermal sphere case.}
\end{figure}

\subsection{Constraints based on the simulation data}
\begin{figure}
\centering
\includegraphics[width=8.5cm]{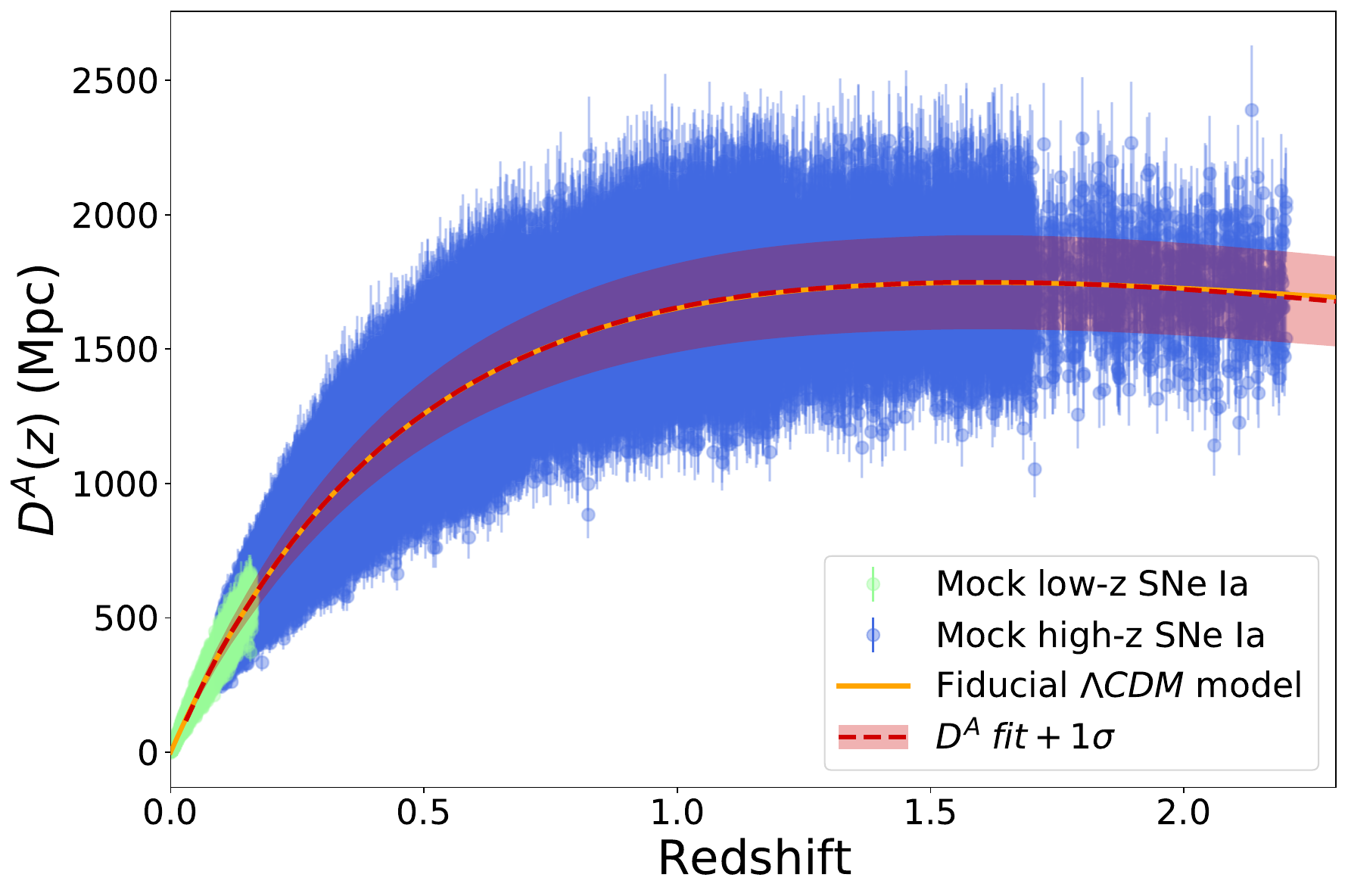}
\caption{\label{fig:sim reconstruction}Reconstruction of angular diameter distance using mock SNe Ia data. Green and blue dots indicate the angular diameter distance calculated from the low-$z$ and high-$z$ SNe Ia, respectively. Orange solid line indicates the redshift-distance relation based under a flat $\mathrm{\Lambda}$CDM cosmology with $\Omega_{M}=0.3$ and $h=0.7$. Red dashed line represents the best-fitted distance reconstruction obtained by ANN+SNIa. The shaded region shows the corresponding 1$\sigma$ uncertainties.}
\end{figure}

\begin{figure}
\centering
\includegraphics[width=\linewidth]{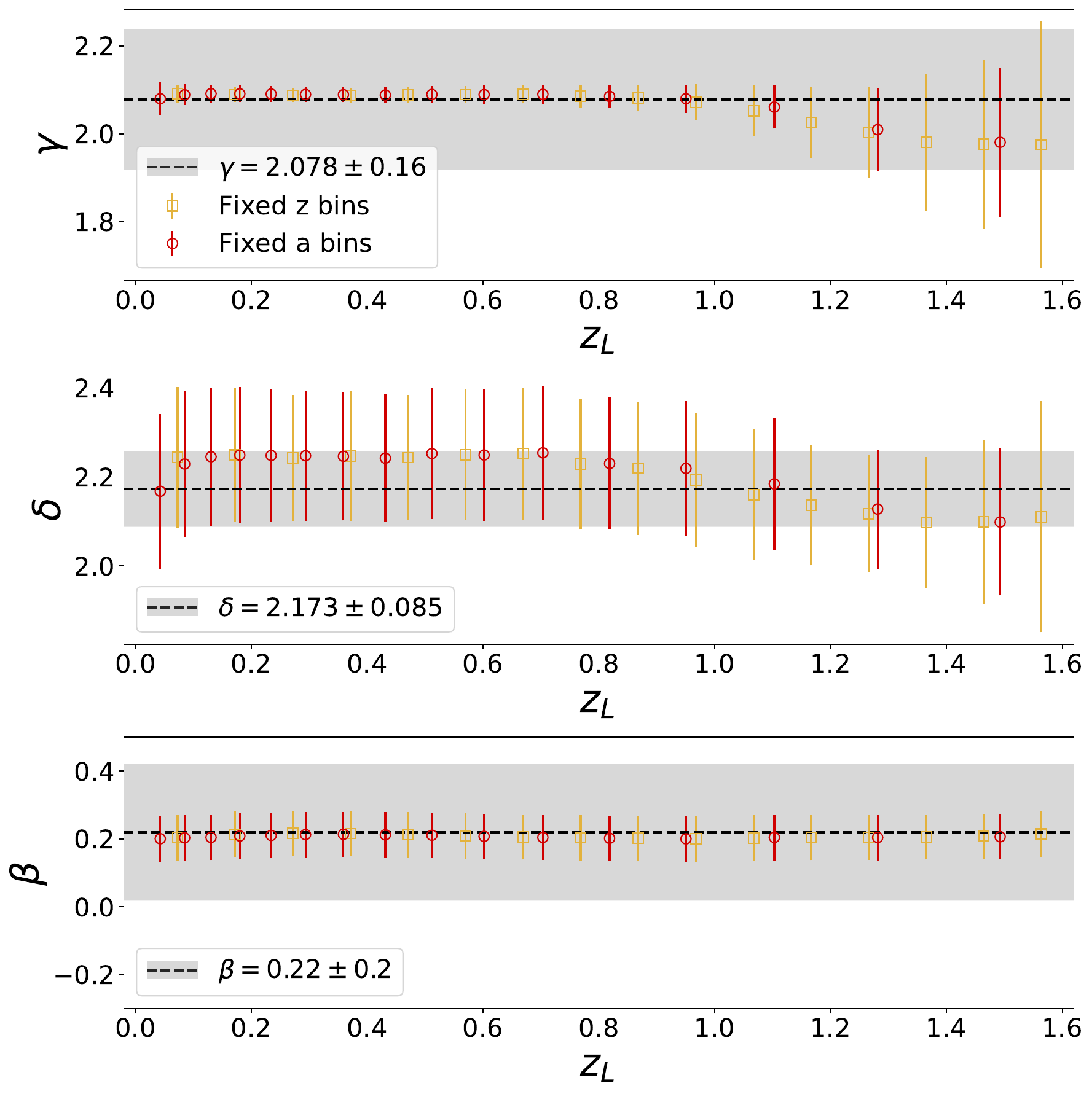}
\caption{\label{fig:sim binning}Constraints on the density slope parameters are shown for redshift bins (hollow orange squares) and scale factor bins (hollow red circles). The gray shaded regions represent the 1$\sigma$ intervals for the fiducial density slope parameters for the simulation.}
\end{figure}

The relationship between redshift and distance reconstructed from mock SNe Ia data is displayed in Fig.~\ref{fig:sim reconstruction}. The ANN effectively reproduces the fiducial redshift-distance function anticipated in the flat $\Lambda$CDM model, demonstrating the precision of our method. 
For the mock SGL data, we employed the individual approach with a single-density slope model, the binning approach and the direct method using an EPL model to assess redshift evolution. The individual approach provided the constraint $\gamma^{sim} = 1.985(\pm 0.003) + z_L \times 0.004(\pm 0.006)$, indicating a nearly constant density slope. 
However, the mean total density slope deviates from the fiducial value due to the mismatch between the  EPL model used to generate the data and the SIS model assumed in the individual analysis.
The results from the binning approach are presented in Figure~\ref{fig:sim binning}. Both redshift and scale-factor binning successfully recover the fiducial values within the 68\% confidence interval, in particular for the total mass density slope at $z \leq 1$. For $z > 1$, some deviations in $\gamma$ and $\delta$ are observed, which are further discussed in Section~\ref{sec:discussion_bias}. In comparison, the anisotropy parameter $\beta$ demonstrates greater robustness than the mass density slope parameters.

\begin{figure}
\centering
\includegraphics[width=\linewidth]{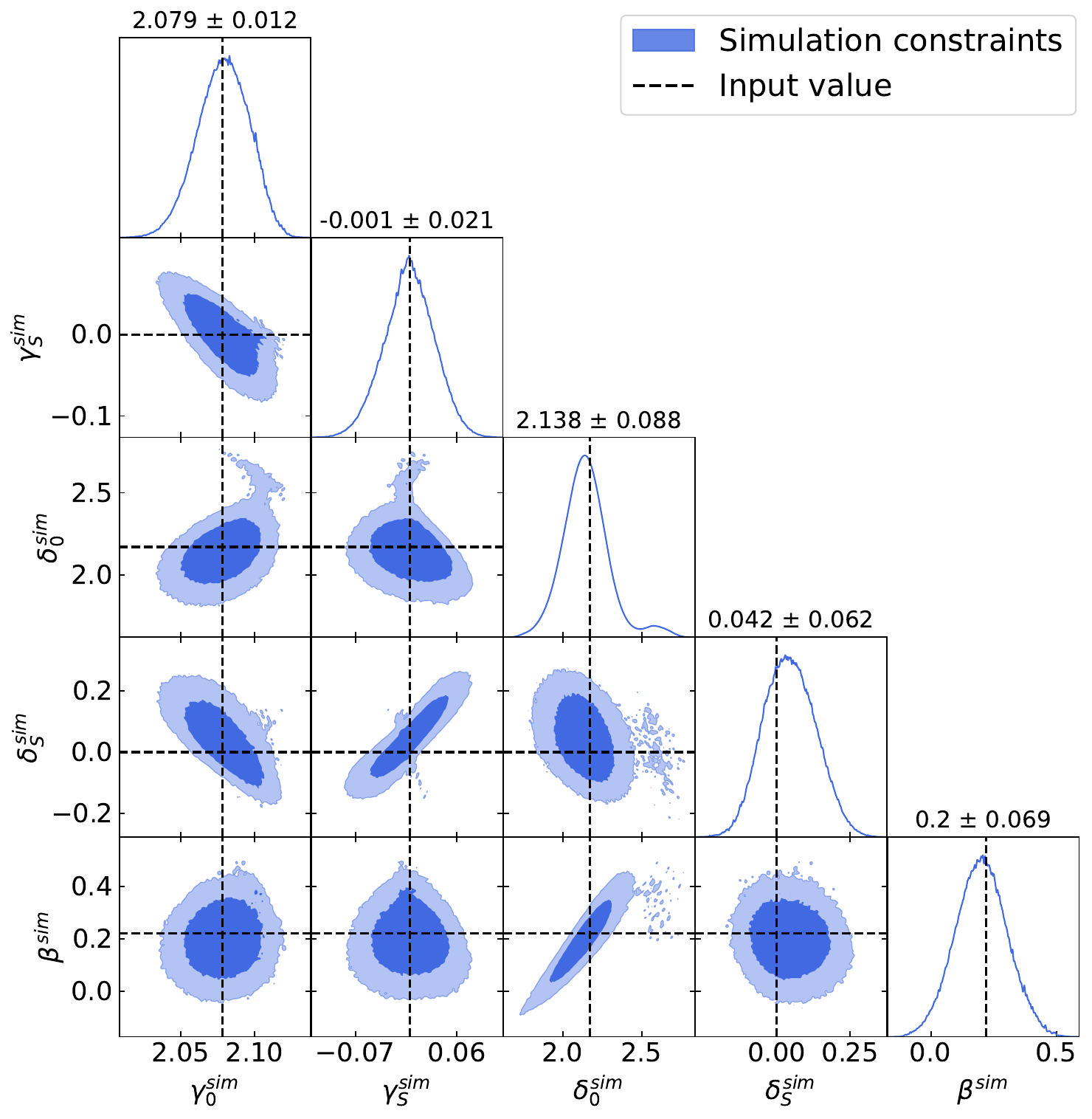}
\caption{\label{fig:sim z evo}Posterior distributions of the spherical power-law density model parameters for 7,506 SGL mock data. Black dashed lines indicate the fiducial values used to generate the mock data.}
\end{figure}

Direct constraints on density slope parameters from the mock data are shown in Figure~\ref{fig:sim z evo}. These parameters are also recovered within a 68\% confidence level. Compared to the parameters related to luminosity matter density and anisotropy, the total mass density parameters ($\gamma_0$ and $\gamma_s$) are constrained with higher precision. The simulation results reveal significant degeneracies, similar to those observed in real data, particularly between the parameters $\delta_0$ and $\beta$, as well as $\gamma_s$ and $\delta_s$. These correlations mainly result from the physical interaction between the luminous tracer and the total mass distribution. High-precision dynamical measurements, such as those provided by integral field units (IFUs), have the potential to break the degeneracy between $\delta_0$ and $\beta$, thereby offering tighter constraints on the evolution parameter $\delta_s$ for luminous tracers. Unlike the individual and binning methods, the direct constraining approach shows reduced sensitivity to outliers in the data, making it a robust alternative for parameter estimation in the presence of anomalous data points.

\section{Discussions}\label{sec:discussions}

In this section, we explore several aspects of our methodology and findings. In Section \ref{sec:discussion_dis_recon}, we discuss how non-parametric methods and model-independent observations are combined to reconstruct distances.
Section \ref{sec:discussion_bias} addresses potential biases in our analysis, while Section \ref{sec:discussion_dependence} considers additional dependencies of the total mass density slope on lens galaxy properties. Finally, in Section \ref{sec:discussion_comparison}, we compare how the derived constraints on the total mass density slope differ across multiple approaches and under varying priors.

\subsection{Reconstructing Distance with Model-independent Observations}\label{sec:discussion_dis_recon}
SGL systems are increasingly recognized as valuable cosmological probes. While their capability for constraining the Hubble constant ($H_0$) through time-delay cosmography is well-established, SGL systems can also be harnessed to investigate additional cosmological parameters, including the matter density ($\Omega_m$), cosmic curvature ($\Omega_k$), and the dark energy equation of state ($w$). Their sensitivity to these parameters has been well-documented \citep{suyu2014cosmology,cao2015cosmology,arendse2019lowredshift,liu2019implications,liao2019modelindependent,wong2020h0licowa,liao2020determining,qi2021measurements,li2024cosmology}, yet it is often not adequately taken into account in current lens modeling approaches. Conventional modeling typically relies on a fiducial cosmological model to calculate distances at specific redshifts, potentially introducing systematic biases. To mitigate these biases and avoid circular reasoning, we constrain the lens model and its redshift evolution solely using observational distance indicators, without assuming any specific cosmological parameters in the flat $\mathrm{\Lambda}$CDM framework. This method, however, only provides discrete measurements at certain redshifts, requiring a suitable reconstruction of the redshift–distance relation to properly match each SGL system.

Our analysis uses ANNs, a non-parametric method, to reconstruct the angular diameter distance at various redshifts. ANN is proven effective in smoothing and accurately reconstructing the redshift-distance relationship, as demonstrated in Fig.~\ref{fig:sim reconstruction} by our simulations trained with extensive SNe Ia data. The method also shows capability in extrapolating beyond the limits of the existing data. However, the precision of the reconstruction is limited by the sample size; a smaller dataset can reduce the accuracy of the results.

\subsection{Potential Sources of Bias}\label{sec:discussion_bias}

\begin{figure}
\centering
\includegraphics[width=\linewidth]{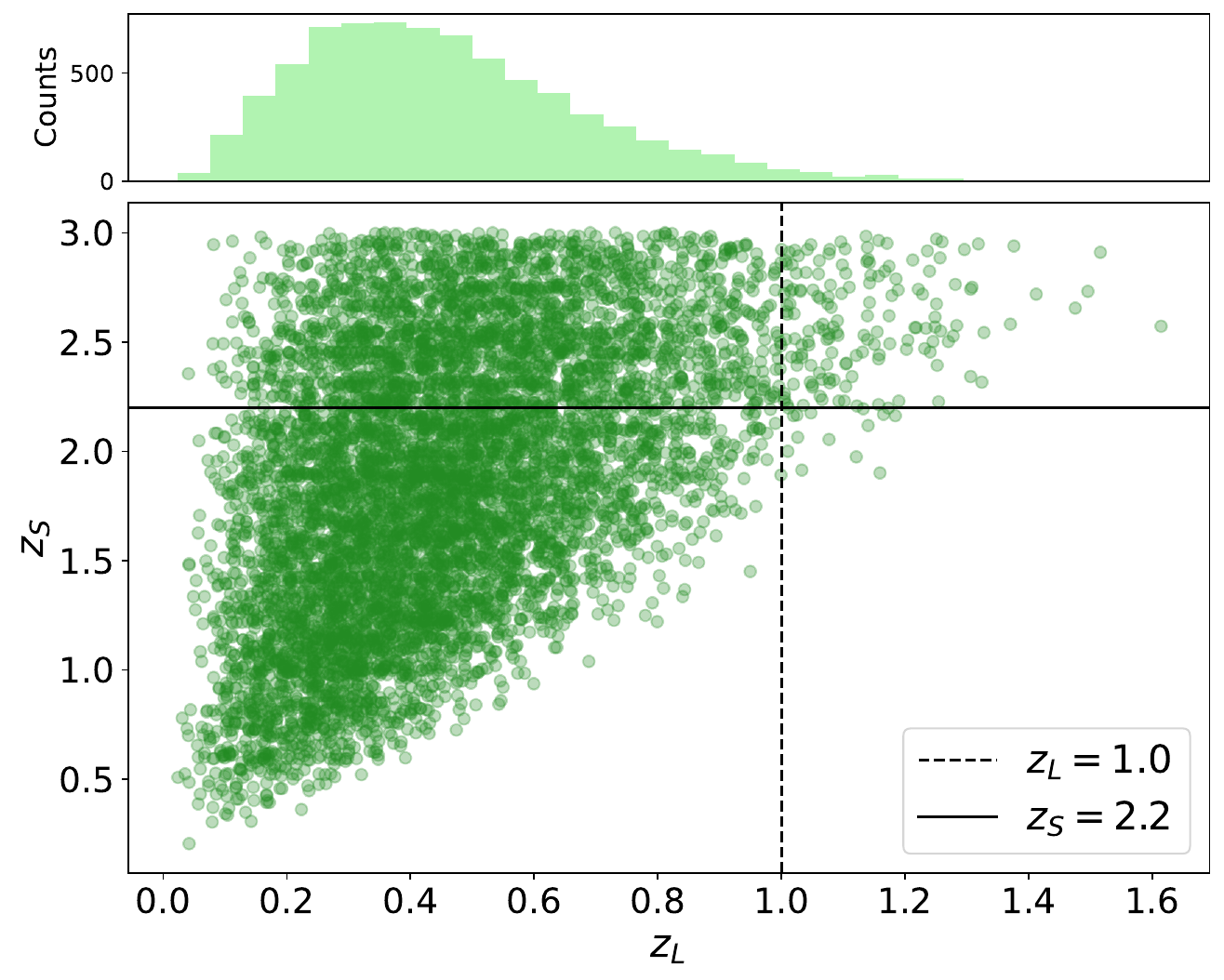}
\caption{\label{fig:sim SGL scatter}The redshift distribution of the simulated SGL. The top histogram shows the redshift distribution of mock lensing galaxies. The dashed black line indicates the lens redshift $z_L=1.0$. The solid black line indicates the source redshift $z_S=2.2$, which is the upper limit for the SNe Ia simulation.}
\end{figure}

When interpreting our findings, it is important to consider potential biases that might affect the accuracy and reliability of results. A significant source of bias may stem from deviations in the ANN reconstruction for D$^A$, especially for data involving high redshifts. As illustrated in Figure~\ref{fig:sim binning}, the total mass density slope $\gamma$ begins to deviate at $z \gtrsim 1$. Two key factors contribute to this deviation.
First, our sample lacks sufficient lensing galaxy data at $z>1$, which is depicted in Fig.~\ref{fig:sim SGL scatter}. The distance ratio, used for constraining density slope parameters, relies on the reconstructed distances from both the lens and the source. 
Without direct observational constraints above \(z = 2.2\), these source distances must be extrapolated from the ANN-based reconstruction. Second, any deviation from the fiducial cosmology in this extrapolation, especially at \(z > 2.2\), can further skew the inferred density slope parameters.
These considerations underscore the importance of careful treatment in future analyses. Insufficient sample sizes and uncertainties in high-redshift extrapolations may similarly affect density slope measurements in real observational datasets, highlighting the need for more comprehensive data coverage at the high-redshift end.

Another potential source of bias could come from the velocity dispersion aperture correction, which may not be appropriate for all measurements. The velocity dispersions ($\sigma_{ap}$) in our sample are measured within different fiber apertures, while the lensing galaxies have different angular sizes spanning a wide dynamic range. To normalize the measured velocity dispersions to a physically meaningful radius, which is half of the effective radius ($\theta_{eff}$) in this work, we apply an aperture correction. We use the power-law function proposed by \citet{jorgensen1995spectroscopy} to quantify the power-law slopes of the normalized $\sigma_{ap}$ profiles (see Eq.~\ref{eq:aperture correction}). The parameters of the correction function were determined by \citet{cappellari2006sauron} based on the SAURON IFS data from 40 ETGs. However, \citet{zhu2023velocity} demonstrated that these parameters may not be applicable to galaxies with intermediate Sersic indices, and the power-law index of the correction function strongly depends on the r-band absolute magnitude $M_r$, $g-i$ color, and Sersic index $n_{ser}$. In this work, we lack the complete information needed to customize the aperture correction for each lensing system, and the changes in the aperture correction can lead to additional bias. A detailed study on applying customized aperture corrections for the lensing galaxies is in preparation.

Additionally, the simplicity of the lens model adopted in our study may introduce additional biases in the constraints. We assume that both the total mass and the luminous tracer in all strong gravitational lensing (SGL) systems follow a spherical power-law profile. Moreover, while anisotropy can vary across different galaxies, our analysis treats it as a universal value for all systems, applying only a broad prior. These assumptions are strong and may not hold in all cases. Indeed, deviations from the power-law model, particularly in the stellar mass distribution, have been reported in studies such as \citet{etherington2022bulgehalo,tan2023project}. Although the power-law model is proven to be valid at the population level, it may deviate from the true value when extreme systems are present in a relatively small subsample.

\subsection{Other Possible Dependence of the Evolution in Density Slope constraints}\label{sec:discussion_dependence}
In analyzing a non-evolving EPL model, we detect bimodal characteristics in the posterior distribution of the total mass density slope ($\gamma$), with a primary peak at $\gamma=1.987$ and a secondary slightly steeper peak. This suggests potential internal divergence or temporal variation in $\gamma$. Our study primarily investigates the redshift evolution of $\gamma$, aligning with findings from other studies, but also acknowledges the significance of internal divergence. \citet{etherington2022bulgehalo} explore additional variables such as normalized Einstein radius $\theta_E/{\theta_{eff}}$ and velocity dispersions, which are motivated by lensing and dynamics \citep{sonnenfeld2012evidence,auger2010sloanb,li2018stronglensing}.

\citet{cao2016limits} document divergence in $\gamma$ with respect to velocity dispersion by categorizing galaxies into low, intermediate, and high mass based on their velocity dispersion measurements. They noted a decrease in $\gamma$ with increasing velocity dispersion in a single density slope model, but no such trend when $\gamma \neq \delta$. Conversely, \citet{etherington2022bulgehalo} observed an increase in $\gamma$ at higher velocity dispersion. In our work, we adopt a similar approach with \citet{etherington2022bulgehalo}, examining the linear relationship between velocity dispersion and $\gamma$ for both total and luminous mass. With a truncated triangular $\beta$ prior mentioned in Section \ref{subsubsec:beta_prior}, our fitting yields the relation $\gamma({\sigma_v})=1.22(\pm0.13)+0.59(\pm0.13)\times \mathrm{log_{10}}(\sigma_{v})$, supporting the results from \citet{etherington2022bulgehalo}. For the dependence on the radius, \citet{li2018stronglensing} found that the total mass density slope has an increasing trend with normalized Einstein radii, while \citet{etherington2022bulgehalo} reported a slightly decreasing trend. In this work, We also assess the impact of the normalized Einstein radius on $\gamma$ and find a linear relation as $\gamma({\theta_{E,nor}})=1.859(\pm 0.053)+0.033(\pm0.033)\times(\theta_E/{\theta_{eff}})$, which is consist with the work using the BELLS, GALLERY, and SL2S samples in \citet{li2018stronglensing}.

\begin{figure}
\centering
\includegraphics[width=\linewidth]{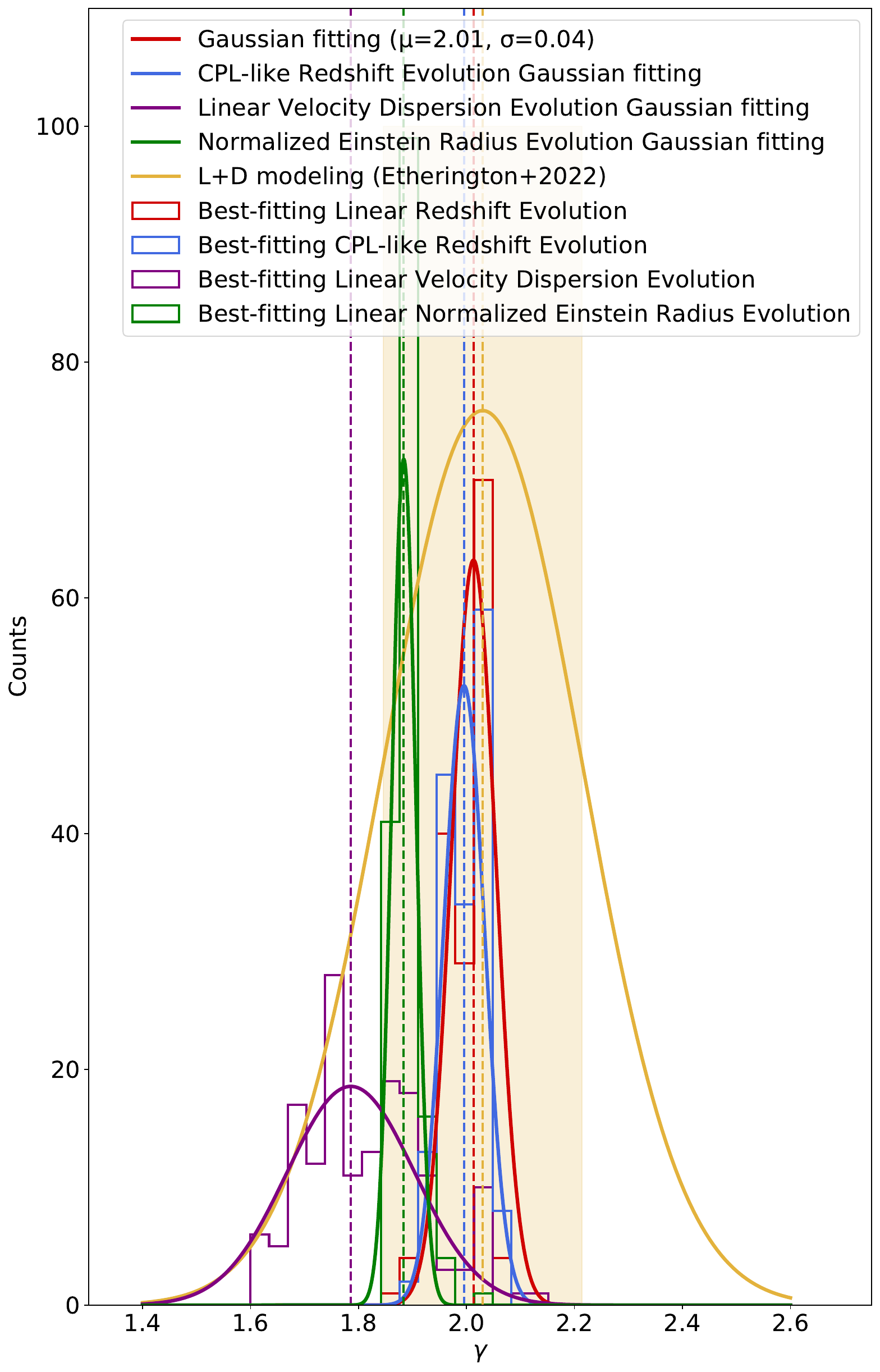}
\caption{\label{fig:evo hist comparison} Regenerated $\gamma$ distributions from various evolutionary models. Red, blue, purple, and green steps represent the distributions from best-fitting linear redshift, ATE, linear velocity dispersion, and linear normalized Einstein radius evolution models, respectively. The solid yellow line indicates the $\gamma$ distribution from lensing and dynamics modeling \citep{etherington2022bulgehalo}, with the shaded area representing the 1 $\sigma$ uncertainty. Dashed lines across all colors mark the mean values from Gaussian fits of each distribution.}
\end{figure}

In order to check if our results can be used as an effective model of population of lenses, we infer the $\gamma$ by using redshifts of lensing galaxies and our best-fitting evolution models under triangular $\beta$ priors. Then we compare the resulting distribution of model-predicted $\gamma$ exponents with the results of detailed modeling \citep{etherington2022bulgehalo} based on observations from 123 lensing galaxies across the SLACS, BELLS, GALLARY, SL2S, and LSD surveys. For comparison we also present the predictions coming from $\gamma({\sigma_v})$ and $\gamma({\theta_{E,nor}})$ regression models. As shown in Fig.~\ref{fig:evo hist comparison}, the $\gamma$ populations derived from linear evolution models based on velocity dispersion and normalized Einstein radii tend to concentrate around lower $\gamma$ values, while our sample of generated $\gamma$ indices closely align within the 1 $\sigma$ uncertainty of the lensing+dynamics modeling results. This consistency supports the validity of the redshift-evolving extended power-law model for describing lensing galaxies at the population level. 

\subsection{Comparing constraints Across Different Approaches and Priors}\label{sec:discussion_comparison}
\begin{figure}
\centering
\includegraphics[width=\linewidth]{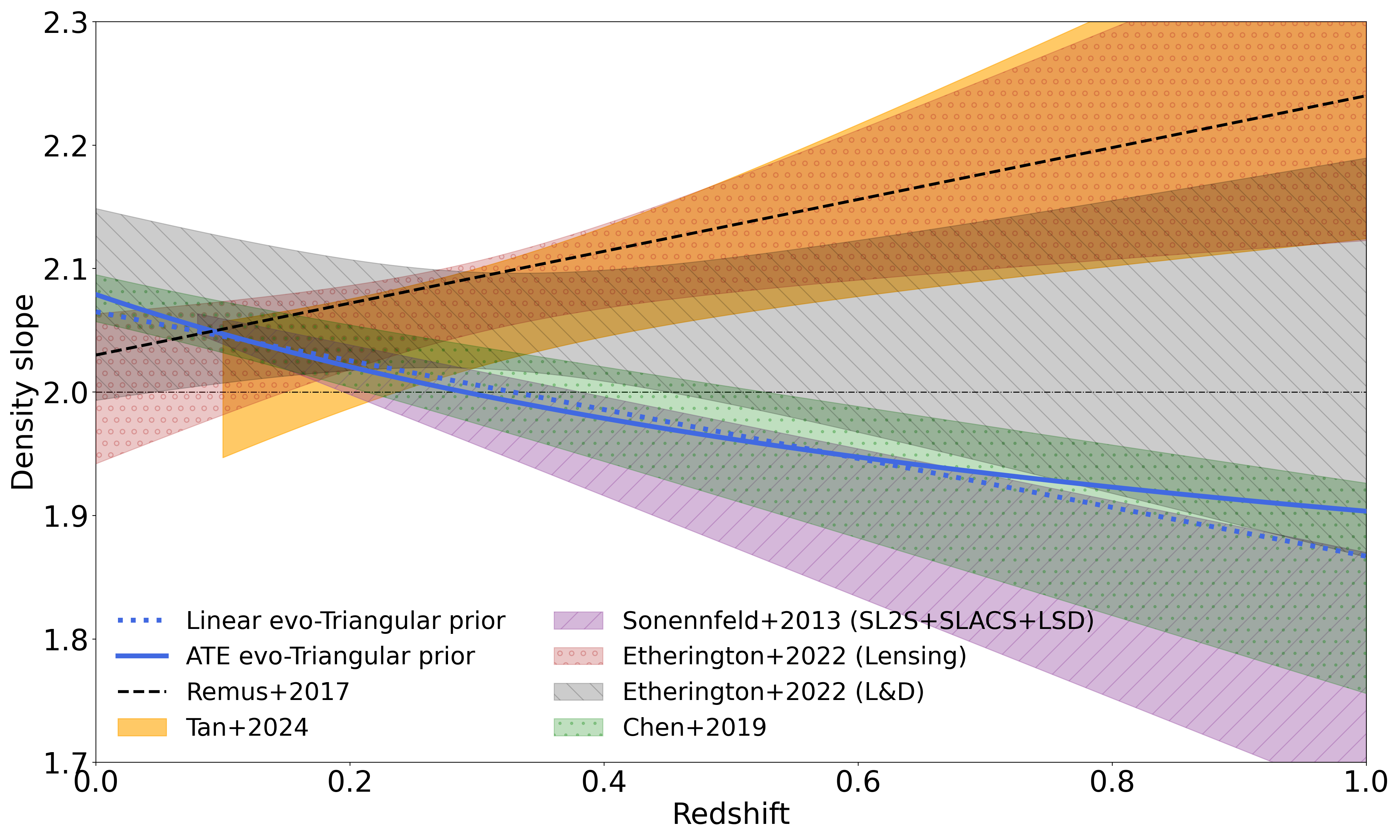}
\caption{\label{fig:z-evo comparison} The total density slope of early-type galaxies as a function of redshift. The blue dotted line represents the best fit using the linear evolution model with a triangular $\beta$ prior, while the blue solid line represents the best fit using the ATE evolution model with a triangular prior described in Section \ref{subsubsec:beta_prior}. The dark green dashed line indicates the average total density slope from the Magneticum simulations \citep{remus2017coevolution}. The light orange band shows the 1 $\sigma$ uncertainty region of constraints on elliptical galaxy mass profile evolution from \citet{tan2024joint}. The purple diagonal striped band represents the total density constraints based on lensing and dynamic information from the SL2S, SLACS, and LSD surveys. The red circled band shows the total density slope evolution constraints obtained by fitting the source light in every pixel of Hubble Space Telescope images from \citet{etherington2022bulgehalo}. The grey backward striped band indicates the lensing+dynamics constraints for overlapping systems with pixelized light fitting constraints from \citet{etherington2022bulgehalo}. The green dotted band represents the single density slope evolution constraints with the same sample used in this paper from \citet{chen2019assessing}.}
\end{figure}
Across various methods, the redshift evolution of the total mass density slope ($\gamma$) of the lensing galaxy consistently indicates either a slightly negative trend or no significant change with increasing redshift. In the case of single density slope, the individual constraints have smallest uncertainties, but with flatter evolution slopes than other two methods. 

In the case of $\gamma\neq\delta$, we adopt three different priors for the anisotropy ($\beta$) of the velocity dispersion for lensing galaxies. Based on 2597 early-type galaxies analyzed using the axisymmetric Jeans Anisotropic Modeling (JAM) method assuming a general NFW dark matter distribution, the triangular prior describes the anisotropy distribution over a broad range. The Gaussian prior, based on the same sample but with additional selection, provides a more consistent and concentrate anisotropy distribution. We find that the differences among the priors do not significantly affect the constraints on the total mass density, but do impact the luminous matter density slope and its redshift evolution. To describe the redshift evolution of the total mass density slope, we consider two semi-analytical evolution models for the direct approach. The linear model is straightforward and widely used to search for evolving trends, while the ATE evolving scenario has a physical basis and is applicable for cases with higher redshifts.

Compared to other studies \citep{sonnenfeld2013sl2sb,remus2017coevolution,chen2019assessing,etherington2022bulgehalo,tan2024joint} on the same topic in Fig.~\ref{fig:z-evo comparison}, our results support a slightly negative or no-evolution scenario for the total mass density slope as redshift increases. The evolution trend is consistent with work based on lensing and dynamic measurements \citep{sonnenfeld2013sl2sb,cao2016limits,holanda2017constraints,chen2019assessing}. However, redshift evolution constraints based on only-lensing data prefer a positive evolution picture \citep{etherington2022bulgehalo,tan2024joint}. This is supported by Magneticum Pathfinder1 simulations, which suggest that gas-poor mergers dominate the evolution phase at lower redshifts \citep{remus2017coevolution}. \citet{etherington2022bulgehalo} points out that these differences may arise because the physical meaning of the density slope differs between lensing+dynamics and only-lensing methods. They argue that the lensing+dynamics method tends to measure the mass-weighted density slope within the effective radius, while the only-lensing method measures the local density slope around the Einstein radius.

For the redshift range $z>0.4$, we see an apparent deviation in the total density slope between results from lensing+dynamics and only-lensing methods in Fig.~\ref{fig:z-evo comparison}. The only-lensing results prefer steeper slopes, while the lensing+dynamics results favor slopes flatter than the isothermal case. This difference in constraints on redshift evolution suggests that the SGL sample is not large enough and is still easily affected by local uncertain data points. It should also be noted that the lensing systems used to constrain redshift evolution differ, which may affect the results due to selection effects. In order to avoid the selection effect,  \citet{etherington2022bulgehalo} compared an overlapping sample with total mass density slope constraints from both lensing+dynamics and only-lensing methods. They found that results from lensing+dynamics prefer an almost no-evolution scenario while only-lensing results support a positive redshift evolution as redshift increases.

\section{Summary}\label{sec:summary}

This study presents a novel dark energy model-independent approach for reconstructing the distance ratios of SGL systems to constrain the mass-density slope and its redshift evolution. To the best of our knowledge, this is the first application of such a method in this field.
To achieve this, we employed a non-parametric ANN method. Using observations of SNIa, we reconstructed distance ratios and compared them to theoretical predictions. This analysis allowed us to estimate the redshift evolution of the  EPL parameters for a sample of 161 lensing galaxies from the LSD, SL2S, SLACS, S4TM, BELLS, and BELLS GALLERY surveys. Mock data, generated using future survey forecasts, was employed to evaluate the ability of our method to accurately recover the input fiducial value and to forecast its constraining power.
Based on our results, we draw the following conclusions:

1. The internal divergence in non-evolving analysis, as observed in Section \ref{res:non-evo} highlights the need to account for redshift evolution in density slopes when analyzing lensing galaxies. Employing a triangular prior on the velocity dispersion anisotropy $\beta$, based on recent MaNGA DynPop dynamical modeling \citep{zhu2024manga}, and a linear redshift evolution model, our best-fitting model indicates redshift evolution as $\gamma = 2.065(\pm0.046) - 0.20(\pm0.12) \times {z}$ for the total mass, and $\delta = 2.14(\pm0.16) - 0.09(\pm0.19) \times {z})$ for the luminous matter.

2. Although observational evidence indicates that the extended power-law (EPL) may not suit individual lens modeling \citep{treu2006sloanh,humphrey2010slope,etherington2022bulgehalo,tan2023project}, we find it remains valid at the population level. This finding supports the use of the ‘bulge-halo conspiracy’ to reduce degrees of freedom in statistically constraining cosmological parameters. As discussed in Section \ref{sec:discussion_dependence}, we observe that the EPL parameters exhibit stronger evolution with redshift compared to other physical quantities.

3. The deviation across different studies may imply that the current sample of lensing galaxies remains too small to produce consistent and robust results regarding the evolution of lenses. This highlights the urgent need for large surveys to uncover new SGL systems and build a more complete dataset.
Wide-area, high-sensitivity optical surveys like the LSST from the Vera C. Rubin Observatory, along with complementary spectroscopic follow-ups using instruments such as 4MOST \citep{depagne20154most} and advanced techniques like IFU \citep{zhu2024manga}, will provide us with a larger and more comprehensive sample\citep{collett2015population}.
The simulation results indicate that increasing the SGL sample size by a factor of fifty could enhance the precision of the total mass density slope evolution to $\Delta (\partial\gamma/\partial z) \sim 0.021$. This improvement would also aid in resolving discrepancies between Lensing+Dynamics and Only-Lensing constraints on the redshift evolution of the total mass distribution in early-type galaxies \citep{etherington2022bulgehalo}. Furthermore, it would provide deeper insights into the differences between gravitational and dynamical mass measurements of lensing galaxies.

\begin{acknowledgements}
      This research was supported by the Polish National Science Center grant 2023/50/A/ST9/00579 and the program of China Scholarships Council. Thanks for the referee for their constructive suggestions to help with this manuscript.
\end{acknowledgements}

\bibliographystyle{aa}
\bibliography{main} 

\begin{appendix}
\section{Complete EPL model constraints}
The appendix presents the complete results of the EPL model parameter constraints, including those for total mass density, luminous mass density, and anisotropy. These results are shown for both the binning approach (see Table~\ref{tab:evo_3D}) and the direct approach (see Table~\ref{tab:evo_5D}).
\begin{table*}[ht]
\centering
\small
\renewcommand{\arraystretch}{1.2} 
\caption{Summary of binning approach results for total mass density, luminous mass density, and anisotropy.}
\label{tab:evo_3D}
\begin{tabular}{c|c|c|c}
\hline \hline
Approach &  $\gamma^{\rm EPL-bin}$ & $\delta^{\rm EPL-bin}$ & $\beta^{\rm EPL-bin}$ \\
\hline
 Fixed z bins + Triangular prior & $\rm\gamma_0 =2.129\pm0.030$  & $\rm\delta_0=1.989\pm0.093$ & $\rm\beta_0 =-0.118\pm0.071$   \\
 Tri(-0.5, 0.656, mode=0.102)    & $\rm\gamma_s =-0.400\pm0.073$ & $\rm\delta_s=0.04\pm0.16$   & $\rm\beta_s =0.22\pm0.12$  \\
\hline
 Fixed z bins + Gaussian prior   & $\rm\gamma_0 =2.119\pm0.024$  & $\rm\delta_0 =2.169\pm0.078$ & $\rm\beta_0 =0.097\pm0.063$   \\
 $N(\mu=0.22, \sigma^2=0.2^2)$   & $\rm\gamma_s =-0.391\pm0.069$ & $\rm\delta_s =-0.11\pm0.13$  & $\rm\beta_s =0.112\pm0.099$  \\
\hline
 Fixed z bins + Gaussian prior   & $\rm\gamma_0 =2.100\pm0.019$  & $\rm\delta_0 =2.243\pm0.063$ & $\rm\beta_0 =0.131\pm0.042$   \\
 $N(\mu=0.18, \sigma^2=0.13^2)$  & $\rm\gamma_s =-0.326\pm0.064$ & $\rm\delta_s =-0.30\pm0.12$  & $\rm\beta_s =0.044\pm0.065$  \\
\hline
 Fixed a bins + Triangular prior & $\rm\gamma_0 =2.104\pm0.038$  & $\rm\delta_0 =2.112\pm0.081$ & $\rm\beta_0 =-0.024\pm0.077$   \\
 Tri(-0.5, 0.656, mode=0.102)    & $\rm\gamma_s =-0.32\pm0.10$ & $\rm\delta_s =-0.12\pm0.16$   & $\rm\beta_s =0.12\pm0.15$  \\
\hline
 Fixed a bins + Gaussian prior & $\rm\gamma_0 =2.098\pm0.036$  & $\rm\delta_0 =2.278\pm0.048$ & $\rm\beta_0 =0.203\pm0.031$   \\
 $N(\mu=0.22, \sigma^2=0.2^2)$ & $\rm\gamma_s =-0.31\pm0.10$ & $\rm\delta_s =-0.32\pm0.11$   & $\rm\beta_s =0.010\pm0.061$  \\
\hline
 Fixed a bins + Gaussian prior & $\rm\gamma_0 =2.101\pm0.036$  & $\rm\delta_0 =2.244\pm0.054$ & $\rm\beta_0 =0.149\pm0.040$   \\
 $N(\mu=0.18, \sigma^2=0.13^2)$    & $\rm\gamma_s =-0.31\pm0.10$ & $\rm\delta_s =-0.28\pm0.12$   & $\rm\beta_s =0.022\pm0.080$  \\
\hline \hline
\end{tabular}
\end{table*}

\begin{table*}[ht]
\centering
\small
\renewcommand{\arraystretch}{1.5} 
\caption{Summary of approaches and results for total mass density, luminous mass density, $\beta$ prior, cAIC, and BIC values.}
\label{tab:evo_5D}
\begin{tabular}{c|c|c|c|c}
\hline \hline
 Approach & Mass density parameters & $\beta$ & cAIC & BIC \\
\hline
                                    & $\rm\gamma_0^{\rm EPL-lin}=2.065\pm0.046$ & &  &  \\
 Linear evolution + Triangular prior & $\rm\gamma_s^{\rm EPL-lin}=-0.20\pm0.12$ & $-0.03\pm0.15$ & 195.70 & 209.64 \\
 Tri(-0.5, 0.656, mode=0.102) & $\rm\delta_0^{\rm EPL-lin}=2.14\pm0.16$ & & & \\
                             & $\rm\delta_s^{\rm EPL-lin}=-0.09\pm0.19$ & & & \\
\hline
                                  & $\rm\gamma_0^{\rm EPL-lin}=2.054\pm0.042$ &  &  & \\
 Linear evolution + Gaussian prior & $\rm\gamma_s^{\rm EPL-lin}=-0.19\pm0.11$ & $0.11\pm0.13$ & 198.06 & 212.00 \\
 $N(\mu=0.22, \sigma^2=0.2^2)$ & $\rm\delta_0^{\rm EPL-lin}=2.26\pm0.13$ & & & \\
                              & $\rm\delta_s^{\rm EPL-lin}=-0.16\pm0.18$ & & & \\
\hline
                                 & $\rm\gamma_0^{\rm EPL-lin}=2.052\pm0.041$ &  & & \\
 Linear evolution + Gaussian prior & $\rm\gamma_s^{\rm EPL-lin}=-0.19\pm0.11$ & $0.133\pm0.085$ & 203.60 & 217.54 \\
 $N(\mu=0.18, \sigma^2=0.13^2)$ & $\rm\delta_0^{\rm EPL-lin}=2.270\pm0.098$ & & & \\
                              & $\rm\delta_s^{\rm EPL-lin}=-0.18\pm0.18$ & & & \\
\hline
                                      & $\rm\gamma_0^{\rm EPL-ATE}=2.079\pm0.056$ &  & & \\
 ATE evolution + Triangular prior & $\rm\gamma_s^{\rm EPL-ATE}=-0.35\pm0.23$ & $0.02\pm0.14$ & 196.77 & 210.71 \\
 Tri(-0.5, 0.656, mode=0.102) & $\rm\delta_0^{\rm EPL-ATE}=2.17\pm0.17$ & & & \\
                             & $\rm\delta_s^{\rm EPL-ATE}=-0.25\pm0.40$ & & & \\
\hline
                                   & $\rm\gamma_0^{\rm EPL-ATE}=2.069\pm0.051$ &  & & \\
 ATE evolution + Gaussian prior & $\rm\gamma_s^{\rm EPL-ATE}=-0.341\pm0.23$ & $0.12\pm0.13$ & 208.20 & 222.14 \\
 $N(\mu=0.22, \sigma^2=0.2^2)$ & $\rm\delta_0^{\rm EPL-ATE}=2.29\pm0.14$ & & & \\
                             & $\rm\delta_s^{\rm EPL-ATE}=-0.36\pm0.39$ & & & \\
\hline
                                   & $\rm\gamma_0^{\rm EPL-ATE}=2.068\pm0.051$ &  & & \\
 ATE evolution + Gaussian prior & $\rm\gamma_s^{\rm EPL-ATE}=-0.34\pm0.23$ & $0.136\pm0.087$ & 209.11 & 223.05 \\
$N(\mu=0.18, \sigma^2=0.13^2)$ & $\rm\delta_0^{\rm EPL-ATE}=2.30\pm0.11$ & & & \\
                               & $\rm\delta_s^{\rm EPL-ATE}=-0.34\pm0.39$ & & & \\
\hline \hline
\end{tabular}
\end{table*}

\end{appendix}

\end{document}